\documentclass{aa} 
\usepackage{graphicx}
\usepackage{epstopdf}
\usepackage[varg]{txfonts}
\usepackage[normalem]{ulem}
\usepackage[usenames]{color}
\usepackage{mathrsfs}
\usepackage{natbib}
\usepackage{enumitem}
\usepackage{hyperref}
\usepackage{float}
\usepackage{afterpage}
\usepackage{soul}
\usepackage{xcolor}
\usepackage{placeins}

\usepackage{graphicx}                           % Use pdf, png, jpg, or eps§ with pdflatex; use eps in DVI mode
\usepackage{amssymb}
\usepackage{nccmath}
\usepackage{booktabs}
\usepackage[tableposition=top]{caption} % Spaces the caption properly
\usepackage{multirow}
\usepackage{hyperref}
\usepackage[font=small,labelfont=bf]{caption}
\usepackage{media9}

\usepackage{txfonts}
\usepackage{xcolor}
\usepackage{soul}

\definecolor{valeriia}{rgb}{0.89, 0.2, 0.5}

\newcommand{\Mm}{{\mathrm{\, Mm}}}
\newcommand{\kms}{{\mathrm{\, km \,s^{-1}}}}

\newcommand{\mins}{{\mathrm{\, minutes}}}

\begin{document} 
        
%       \title{Interaction of Coronal Eruption with Remote Prominence: A 2.5D Numerical Study of EUV-wave driven oscillations}

 \title{Numerical study of solar eruption, extreme-ultraviolet wave propagation, and wave-induced prominence dynamics}
        
        \author{V. Liakh\inst{1}
        \and
        R. Keppens\inst{2}}
        
        % \institute{Instituto de Astrof\'{\i}sica de Canarias, E-38205 La Laguna, Tenerife, Spain\\
         \institute{Rosseland Centre for Solar Physics, University of Oslo,\\
                        PO Box 1029, Blindern 0315, Oslo, Norway
                        \email{valeriia.liakh@astro.uio.no}
            \and
            Centre for mathematical Plasma Astrophysics, Department of Mathematics,\\
                        KU Leuven, Celestijnenlaan 200B, 3001 Leuven, Belgium
                }

                %   \institute{Instituto de Astrof\'{\i}sica de Canarias, E-38200 La Laguna, Tenerife, Spain  \label{inst1}\centering
                        %   \and Universidad de La Laguna, Dept. Astrof\'{\i}sica, E-38206 La Laguna, Tenerife, Spain\\  \email{vliakh@iac.es} \label{inst2}\centering}
                
                \date{}

                \titlerunning{Numerical simulations of LAOs}

                \abstract
{Extreme ultraviolet (EUV) waves, frequently produced by eruptions, propagate through the non-uniform magnetic field of the solar corona and interact with distant prominences, inducing their global oscillations. However, the generation, propagation, and interaction of these waves with distant prominences remain poorly understood.}
{We aim to study the influence of an eruptive flux rope (EFR) on a distant prominence by means of extreme-resolution numerical simulations. We cover a domain of a horizontal extent of 1100 Mm, while capturing details down to 130 km using automated grid refinement.}
{We performed a 2.5D numerical experiment using the open-source \texttt{MPI-AMRVAC 3.1} code, modeling an eruption as a 2.5D catastrophe scenario augmented with a distant dipole magnetic field to form a flux rope prominence.}
{Our findings reveal that the EFR becomes unstable and generates a quasi-circular front. The primary front produces a slow secondary front when crossing the equipartition lines where the Alfvén speed is close to the sound speed. The resulting fast (primary) and slow (secondary) EUV waves show different behaviors: the fast EUV wave slightly decelerates as it propagates through the corona, while the slow EUV wave forms a stationary front. The fast EUV wave interacts with the remote prominence, driving both transverse and longitudinal oscillations. Additionally, magnetic reconnection at a null point below the prominence-hosting flux rope is triggered by the fast EUV wave, affecting the flux rope magnetic field and the prominence oscillations.}
{Our study unifies important results of the dynamics of eruptive events and their interactions with distant prominences, including details of (oscillatory) reconnection and chaotic plasmoid dynamics. We demonstrate for the first time the full consequences of remote eruptions on prominence dynamics and clarify the damping mechanisms of prominence oscillations.}
                
                \maketitle
                %
                %-------------------------------------------------------------------
                %
                %-------------------------------------------------------------------
                
                \section{Introduction}\label{sec:introduction}
  %Prominence and driving of the prominence dynamics
Solar prominences are cool dense plasma clouds typically located in magnetic dips high in the solar corona, where they are supported against gravity by magnetic forces.  These structures are highly dynamic; one manifestation of this dynamism is prominence oscillations. Such oscillations are generally classified as small-amplitude oscillations (SAOs) or large-amplitude oscillations (LAOs), based on their velocities, with a threshold of $10\ \text{km/s}$. Additionally, prominence oscillations are classified by the direction of plasma motion relative to the magnetic field as longitudinal and transverse, respectively. Further details on the classification and characteristics of these oscillations can be found in the latest update of the living review by \citet{Arregui:2018spr}.

%Large energy involved and what can provide this energy
Large-amplitude oscillations are expected to contain significant energy due to the high mass and velocities associated with prominences. Most observed LAO events have been linked to energetic disturbances. Numerous studies have reported LAO excitation in filaments induced by Moreton and extreme ultraviolet (EUV) waves \citep[see, e. g.][]{Gilbert:2008apj, Asai:2012apjl, RiuLiu:2013apj, Shen:2014apj1, Shen:2017apj, Xue:2014solpol}. A simultaneous excitation of different LAO polarizations (with the polarization identified by the direction of plasma motions) from a single energetic event has also been observed. For example, \citet{Gilbert:2008apj} reported an instance in which the filament oscillations induced by a Moreton wave exhibited mixed polarization modes. Recently, studies by \citet{Devi:2022adsr}, \citet{ZhangY:2024apj}, and \citet{ZhangQM:2024mnras} reported EUV waves from eruption events exciting prominence oscillations. However, the specific mechanisms underlying the generation, propagation, and substantial energy transfer from these disturbances to prominences, which trigger LAOs, remain unclear.

%Histrorical context, Moreton and EIT waves
Early evidence of such waves was reported by \citet{Moreton:1960pasp}, who observed them emanating from flares; these waves propagated through the solar atmosphere at velocities of $500-2000\ \text{km/s}$. The Moreton wave is widely recognized as a fast-mode wave propagating through the chromosphere and low corona. The EUV Imaging Telescope (EIT) aboard the Solar and Heliospheric Observatory (SOHO) detected another wave-like phenomenon in the solar corona, named after the telescope as the EIT wave \citep{Moses:1997solphys, Thompson:1998grl}. These EIT waves were typically associated with coronal mass ejections (CMEs) or eruptions \citep{Biesecker:2002apj, Chen:2006apj, Chen:2011spr}.  Initially, EIT waves were considered the coronal counterparts of Moreton waves \citep{Moses:1997solphys, Thompson:1998grl}. However, this interpretation was challenged by the observed velocities of EIT waves, which were typically about three times lower than those of Moreton waves. Moreover, observations revealed that EIT fronts could become stationary: they were trapped in magnetic arcades, decelerated, and eventually halted near quasi-separatrix layers (QSLs) \citep{Delannee:1999solphys, Delannee:2000apj}. Since fast-mode waves are expected to propagate through QSLs, this behavior implies that EIT waves have a distinct physical nature. Several theories have been proposed to explain EIT waves, including the magnetic field line stretching model \citep{Chen:2002apjl, Chen:2005apj}, the successive reconnection model \citep{Attrill:2007apjl, Attrill:2007asn}, the slow-mode solitons model \citep{Wills-Davey:2007apj}, the current shell model \citep{Delannee:2008solphys}, and the slow-mode wave model \citep{Wang:2009apj}.

%Interpretation for EIT(EUV) waves.
%For the EIT wave, \citet{Yang:2010solphys} found that the negative correlation between the velocity of propagation of this phenomenon and the magnetic field strength contradicts the fast magnetoacoustic wave nature but agrees with the field stretching model proposed by \citet{Chen:2002apjl}. 
The launch of the Solar Dynamics Observatory (SDO) with its Atmospheric Imaging Assembly (AIA) marked a new era in the observation of these phenomena. The EIT waves began to be predominantly referred to as EUV waves due to their detection in this spectral range. Observations with SDO/AIA have provided significant insights into these phenomena. However, the uncertainty regarding the physical nature of EUV waves persisted. \citet{Liu:2010apjl} and \citet{Chen:2011apjl} identified both fast EUV waves and non-wave slow EUV phenomena, aligning with the idea that fast EUV waves are the coronal counterparts of Moreton waves, while the slow EUV phenomena result from the stretching of magnetic field lines. Many observations of the interactions of the EUV waves with the different magnetic objects in the solar corona suggested the fast magnetoacoustic wave nature of EUV phenomena. The EUV wave reflection from a coronal hole has been detected in many observations \citep[see, e.g.,][]{Long:2008apjl, Gopalswamy:2009apjl, Kienreich:2013solphys, Yang:2013apj}. Additionally, \citet{Olmedo:2012apj} observed the interaction of an EUV wave with a coronal hole, which resulted in reflection and transmission. Intriguingly, \citet{Chandra:2016apj} observed a fast EUV wave, creating a secondary slow EUV wave near a QSL. This slow EUV wave decelerated and became stationary. Recently, \citet{Zhou:2024nat} reported the detection of magnetohydrodynamic (MHD) wave lensing in the highly ionized and magnetized coronal plasma, using observations from the SDO/AIA. In this observation, the quasi-periodic wave fronts converge at a specific point after traversing a coronal hole, resembling the lensing of electromagnetic waves from a source to a focal point. 
%This MHD wave lensing was accurately reproduced in a numerical simulation.  
%
%Similarly, \citet{Yang:2013apj} detailed the interaction of EUV fronts with both active region loops and coronal holes, producing reflected waves. \citet{Nitta:2013apj} studied large-scale coronal propagating fronts and found that their speeds were not strongly correlated with flare intensities or CME dynamics, aligning more closely with the wave interpretation. 
%
%\citet{Lulic:2013solphys} described the formation of a coronal shock wave driven by an expanding cylindrical piston, noting that the wave detached from the CME after a specific time. 
%Their observations also included rarefaction regions forming dimming areas and compression at CME edges, possibly linked to slow EUV waves.
%Intriguingly, \citet{Chandra:2016apj} observed a fast EUV wave, creating a slow EUV wave near a QSL. This slow EUV wave decelerated and became stationary. 
%Both fast and slow EUV waves have been observed more frequently \citep{Chandra:2018apj, Cunha-Silva:2018aap, Chandra:2018adsr}. 
%Using combined EUV and H$\alpha$ observations, \citet{Wang:2020apj} concluded that Moreton and EUV waves represent distinct phenomena originating from a fast magnetoacoustic shock wave in the corona, propagating through layers associated with different temperatures.
The observations above suggest that fast EUV phenomena are likely fast magnetoacoustic waves, acting as the coronal counterparts of Moreton waves. They are often initiated by eruptions and propagate through the non-uniformly magnetized solar corona. The slow EUV phenomena can be explained as slow magnetoacoustic waves or from the progressive stretching and opening of magnetic field lines during an eruption.

%Numerical simulations of the coronal waves produced by the eruption
Numerous attempts have been made to model Moreton, EUV, and EIT waves using 2D and 3D numerical simulations. \citet{Chen:2002apjl} performed simulations of eruptive flux ropes (EFRs) and studied the origins of Moreton and EIT waves. Their results revealed a piston-like shock surrounding the flux rope, with its skirt sweeping the solar surface at speeds exceeding $700\ \text{km/s}$. This shock skirt was identified as the coronal Moreton wave. Additionally, they observed a slower non-wave phenomenon propagating at approximately $250\ \text{km/s}$, which was attributed to an EIT wave formed by the progressive stretching and opening of the magnetic field lines around the EFR. In these simulations, an enhanced density region developed ahead of the stretching field lines forming the EIT wave, while dimmings appeared in the inner region behind it.

Similarly, data-driven simulations have been employed to reproduce coronal wave dynamics \citep{Cohen:2009apj, Downs:2011apj, Downs:2012apj, Downs:2021apj}. \citet{Cohen:2009apj} simulated diffuse bright fronts observed by the Extreme Ultraviolet Imager (EUVI) on board the Solar Terrestrial Relations Observatory (STEREO) and distinguished between wave and non-wave EUV phenomena. \citet{Downs:2011apj} and \citet{Downs:2012apj} detected a fast magnetoacoustic front in their experiments, which became detached from the eruption, and a compression front directly linked to the eruption itself. 

Using a 2.5D catastrophe scenario, \citet{Wang:2009apj} and \citet{Wang:2015apj} identified a fast magnetoacoustic wave, interpreted as a fast EUV wave, along with a slow magnetoacoustic wave, echoes, and vortices, which were suggested to correspond to slow EUV waves. \citet{Mei:2012scpma} showed that stronger background fields and lower densities produced more energetic eruptions. An intriguing feature in their study was the detection of secondary echoes due to the interaction between an echo, a slow-mode shock, and vortices. They associated all these phenomena with slow EUV waves. Using a 3D model, \citet{Mei:2020mnras} confirmed the presence of fast- and slow-mode shocks, echoes, and vortices produced by the eruption, noting the difficulty of the slow-mode shock front detection in synthetic EUV images compared to the fast-mode shock front.

%\citet{Chen:2016solphys} explored the formation of stationary fronts after passing the primary front through QSLs. They concluded that fast-to-slow mode conversion at the QSL resulted in a slow front that stopped at the next separatrix, forming a stationary front. \citet{Downs:2021apj} further highlighted the potential of studying fast EUV wave kinematics as a diagnostic tool for probing the coronal medium.

%In a recent study, \citet{Hu:2024apj} applied the 2.5D catastrophic scenario with varying magnetic field parameters. They observed the generation of multiple fast magnetoacoustic fronts during the quasi-stable phase of the experiment, followed by the gradual rise and eruption of the flux rope, which produced the main fast wavefront. 

%Numerical simulations of the interaction of the eruptions and coronal waves with coronal holes, loops, and prominences or filaments.
From a theoretical perspective, the interaction of Moreton and EUV waves with surrounding magnetic arcades, coronal holes, and prominences has been extensively studied \citep[see, e.g.,][]{Piantschitsch:2017apj, Piantschitsch:2023aap, Piantschitsch:2024aap, Afanasyev:2018aap, Liakh:2020aap, Liakh:2023aap, Zurbriggen:2021solphys}. \citet{Liakh:2020aap} simulated wave triggering using an artificial perturbation to mimic an energetic disturbance, such as a flare, located at a certain distance from the prominences. Building on this, \citet{Liakh:2023aap} employed a more advanced model to study the self-consistent triggering of the prominence caused by the nearby eruption. In this scenario, the triggering mechanism was primarily due to the evolution of plasmoids in the current sheet. Despite the important results obtained from both observational and numerical studies, significant gaps remain in our understanding of EUV waves formation, the propagation in a non-uniformly magnetized coronal medium, and the interaction of these perturbations with distant prominences. In this paper we comprehensively study all these aspects using a 2.5D numerical model.

This paper is organized as follows. In Sect. \ref{sec:numerical} we describe the numerical setup of the experiment. In Sect. \ref{sec:results} we present the main results, describing the different stages of the numerical experiment, including the onset of the eruption, the propagation of the front, its interaction with the prominence, and the response of the prominence to the perturbation.
 Finally, in Sects. \ref{sec:discussion} and \ref{sec:conclusions} we discuss and summarize our main findings.

                \section{Numerical setup}\label{sec:numerical}
     \begin{figure*}[!h]
        \centering
        \includegraphics[width=0.9\textwidth]{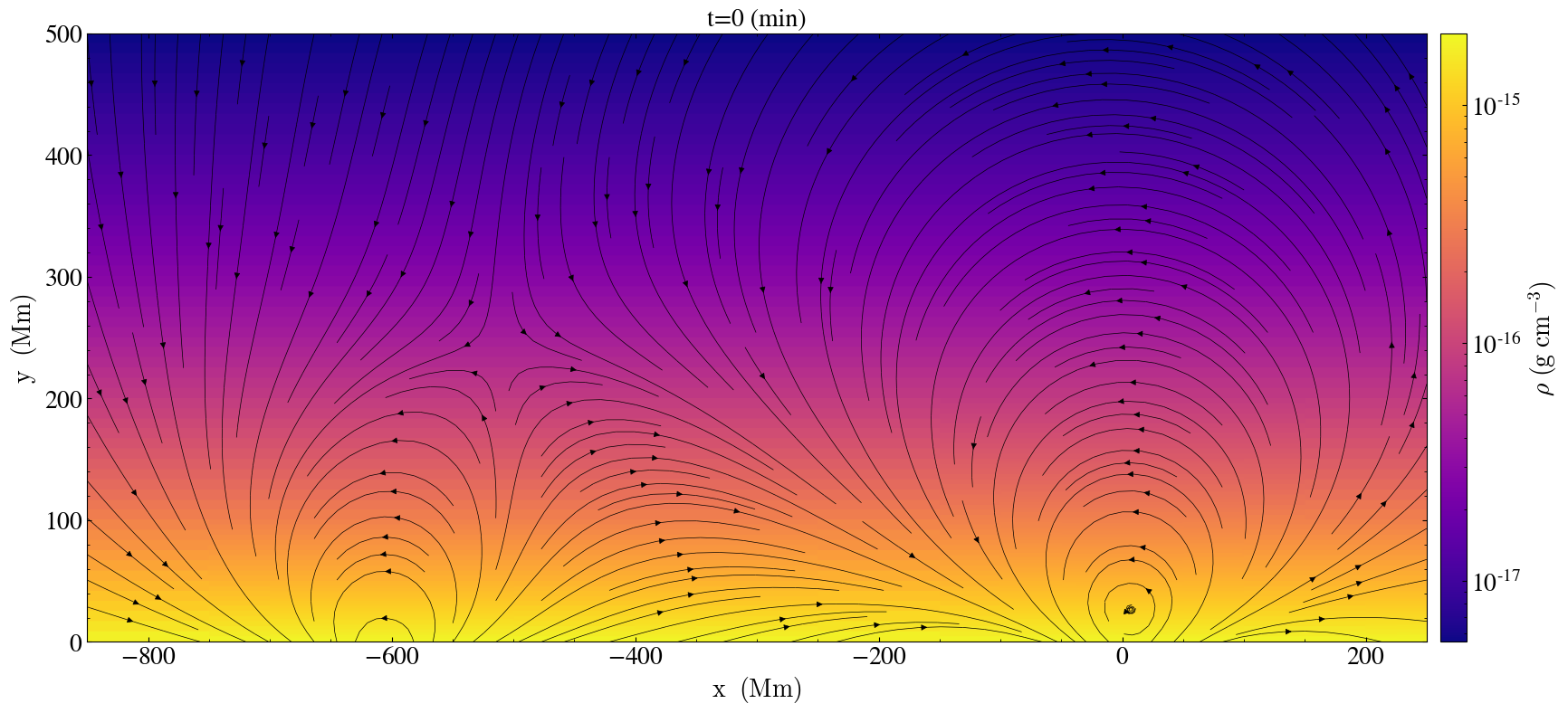}
        \caption{Initial density distribution and magnetic field lines in the entire numerical domain. Animation 1 shows the global evolution of the density, temperature, $v_{\parallel}=(\mathbf{v}\cdot\mathbf{B})/B$, and $v_{\perp}=v_y-v_{\parallel} B_y/B$ during the entire simulation time. An animation of this figure is available online.  \label{fig:setup}}
     \end{figure*}  
     \begin{figure}[!t]
        \includegraphics[width=0.45\textwidth]{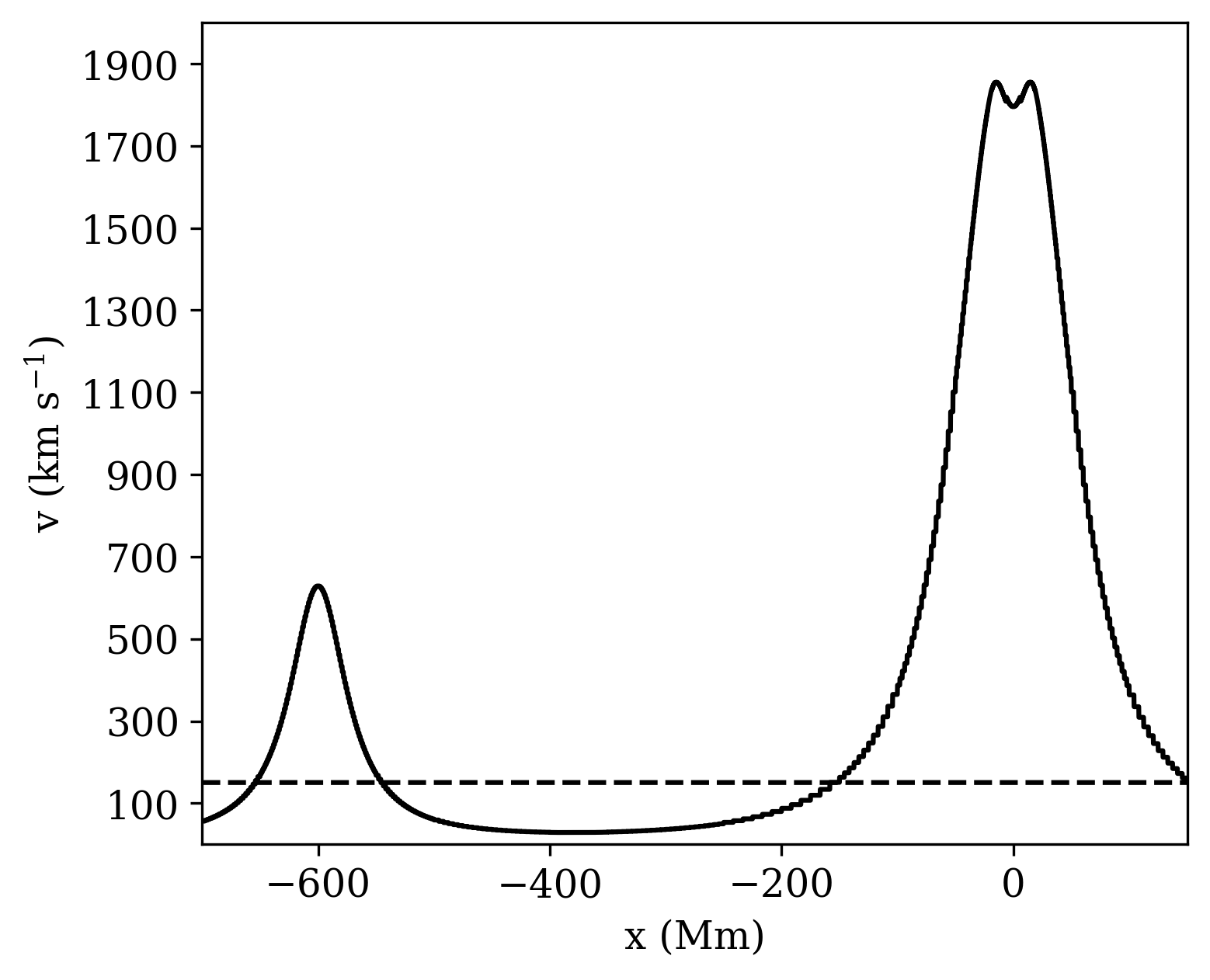}
        \includegraphics[width=0.45\textwidth]{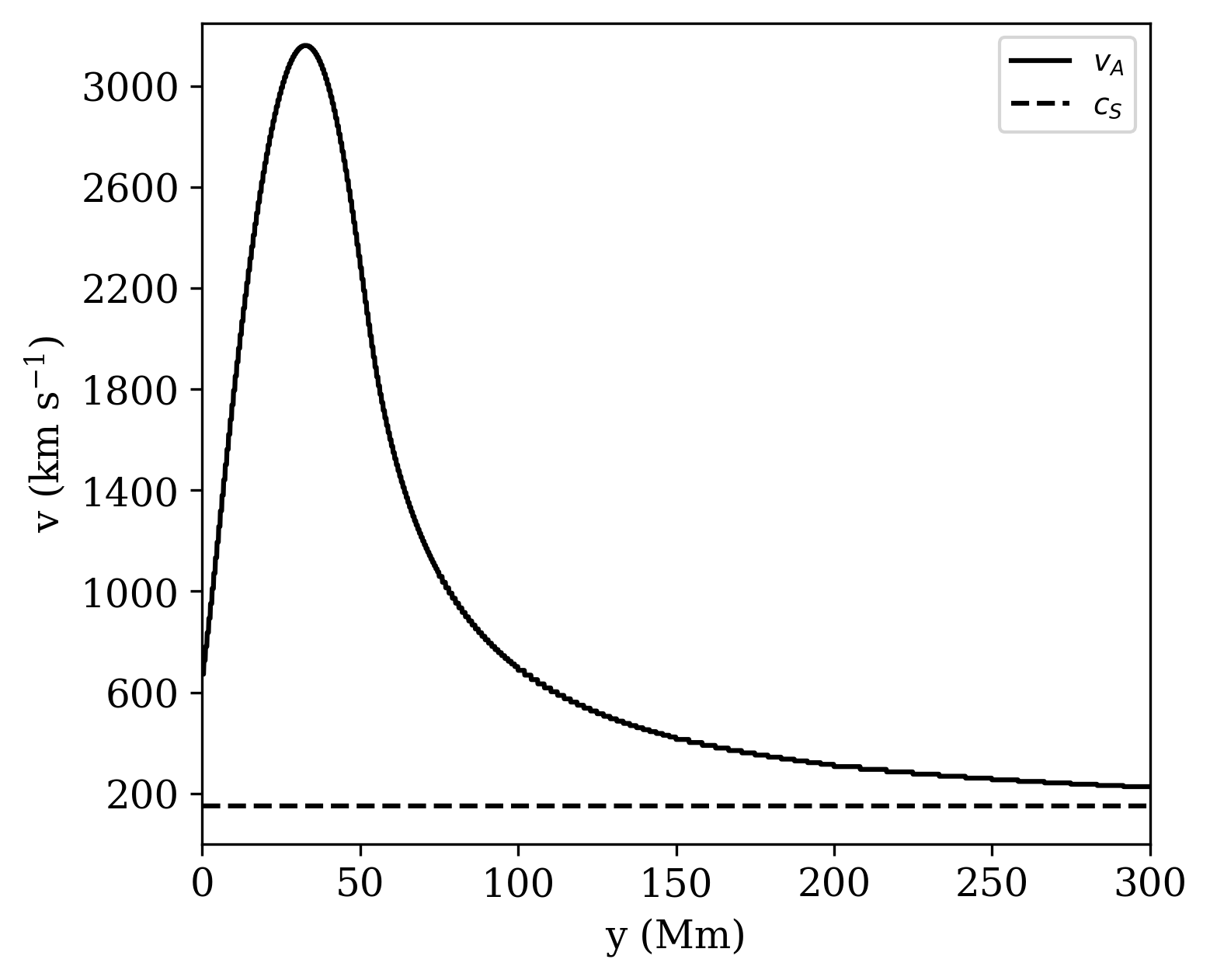}
        \caption{Alfvén and sound speed along the horizontal cut at $y=10$ Mm (top) and the vertical cut at $x=0$ Mm (bottom). \label{fig:speeds}}
     \end{figure} 
 
The numerical experiment was performed using the fully open-source, adaptive-grid, parallelized Adaptive Mesh Refinement Versatile Advection Code (\texttt{MPI-AMRVAC 3.1})\footnote{\texttt{MPI-AMRVAC 3.1}, available at \href{http://amrvac.org}{http://amrvac.org}.} \citep{Porth:2014apjs, Xia:2018apjs, Keppens:2021cmwa, Keppens:2023aap}.
We used a Cartesian coordinate system, with the $x$- and $y$-axes denoting the horizontal and vertical directions, respectively. The numerical domain had a physical size of $1100 \times 500$ Mm and consisted of $132 \times 60$ grid cells at the base resolution. To achieve higher resolution, where needed, we applied seven levels of AMR, resolving structures down to the smallest grid cell size of $130.2$ km. 
This resolution is comparable to that of previous 2.5D numerical studies on wave-induced prominence dynamics \citep{Liakh:2020aap, Zurbriggen:2021solphys, Liakh:2023aap}. We used a Lohner-type AMR prescription \citep{Lohner:1987} with second-order gradient evaluations of density and magnetic field components.
Additionally, the base resolution was enforced in three specific regions near the top and side boundaries: $x < -750$ Mm, $x > 50$ Mm starting at $t = 5.7\mins$, and $y > 300$ Mm throughout the entire numerical experiment.         
                
The \texttt{MPI-AMRVAC 3.1} code solves MHD equations that include non-ideal, non-adiabatic, and various physical source terms, such as the solar gravitational field. The MHD equations can be found in textbooks \citep[e.g.,][]{bookhans} and are equivalent to Eqs. (2-5) as presented in \citet{Brughmans:2022aap}. For the equation of state, we used the ideal gas law for a monoatomic gas with a specific heat ratio $\gamma=5/3$. The mean molecular mass is $\mu\approx0.6$, assuming fully ionized plasma with the Helium abundance $n_{He}=0.1n_{H}$. The energy equation contains terms of optically thin radiation, anisotropic thermal conduction, and Ohmic heating due to a numerical magnetic resistivity. 
%We also add small physical resistivity which is chosen at $\eta=2\times10^{-10}$ in code units ($2.33\times 10^5\ \mathrm{cm^{2}\ s^{-1}}$). 
The energy balance equation also includes a fixed background heating term, an exponential decay with height, to compensate initially for radiative losses.

The set of MHD equations was solved using the Harten–Lax–van Leer flux scheme (HLL) \citep{Harten:1983} with the second-order symmetric Total Variation Diminishing (TVD) slope limiter \citep{Vanleer:1974}. 
The time integration was performed using the Runge-Kutta three-step method. 
                %
%In order to ensure $\nabla\cdot\mathbf{B}=0$ condition, we use the parabolic diffusion method \citep{Keppens:2003,Keppens:2023aap}. 
                %
%For the thin radiative losses, we use \textsc{Colgan-DM} cooling curve from \citet{Colgan:2008apj}, but extended with a low-temperature treatment using $12000$ points for the resolution of the temperature in an interpolated table \citep[for details see][]{Hermans2021}. In order to add the optically thin radiative losses, we use the exact integration scheme \citet{Townsend:2009apjs}.  
To ensure the $\nabla\cdot\mathbf{B}=0$ condition, the parabolic diffusion method was applied \citep{Keppens:2003,Keppens:2023aap}.
For optically thin radiative losses, we used the \textsc{Colgan-DM} cooling curve from \citet{Colgan:2008apj}, extended with a low-temperature treatment using $12,000$ points to resolve the temperature in an interpolated table \citep[for details see][]{Hermans2021}. 
The optically thin radiative losses were added using the exact integration scheme from \citet{Townsend:2009apjs}. 

The initial atmosphere is a gravitationally stratified corona with a constant temperature $T_0=1$ MK and the gravitational acceleration defined as $g(y)=g_{\odot}R_{\odot}^2/(R_{\odot}+y)^2$, where $g_{\odot}=2.74\times 10^{4} \mathrm{cm\ s^{-2}}$ is the gravitational acceleration at the solar surface, and $R_{\odot}=696.1\Mm$ is the solar radius. The pressure scale height then varies along the vertical direction as $H(y)=H_{0}(R_{\odot}+y)/R_{\odot}$, where $H_{0}\approx50$ Mm is the pressure scale height at the bottom. The pressure, density, and background heating have the following values at $y=0$: $p_{0,bot}=0.298\ \mathrm{dyn\ cm^{-2}}$, $\rho_{0,bot}=2\times10^{-15}\ \mathrm{g\ cm^{-3}}$, and $\mathcal{H}_b(y)=\rho_0^2\Lambda(T_0)e^{-2y/H(y)}=3.972\times 10^{-4}\ \mathrm{ergs\ cm^{-3}\ s^{-1}}$. 

We included anisotropic thermal conduction $\nabla\cdot(\overleftrightarrow{\kappa}\cdot\nabla T)$ along the magnetic field lines, using the Spitzer conductivity $\kappa_{\parallel}=8\times 10^{-7}\ T^{5/2}\ \mathrm{ergs\ cm^{-1} s^{-1} K^{-1}}$ \citep{spitzer2006physics}. In this experiment, $\kappa_{\perp}$ was neglected due to its comparably small value compared to $\kappa_{\parallel}$. In order to add the parabolic source term of the anisotropic thermal conduction, we used the Runge-Kutta Legendre super-time-stepping technique \citep[RKL;][]{Meyer:2014jorcomsci}.

Figure \ref{fig:setup} presents the magnetic configuration, which consists of the 2.5D catastrophe magnetic field described by \citet{Takahashi:2017apj} and a dipolar magnetic field. The 2.5D catastrophe magnetic field consists of a cylindrical current, an image current beneath the lower boundary to produce an upward magnetic force, and a background quadrupolar magnetic field producing a balancing downward magnetic force. This setup has various parameters setting the rope center and magnetic field strength. Then, the eruption occurs when the parameter of the quadrupolar field strength is $M_q < 27/8$. We set $M_q=0.8\times27/8$ in our configuration in order to obtain an immediate eruption. The EFR was centered at $x=0$ and $y=27$ Mm, with a radius of $R=27$ Mm, and the magnetic field strength at the EFR center, $B=36.6$ G. The 2.5D catastrophe model was chosen because it offers a straightforward and effective method for producing eruptions within a 2.5D numerical setup. 
%In this scenario, the EFR is initially unstable, generating a primary front.
        %
The dipolar magnetic field was centered at $x=-600$ Mm (see Fig. \ref{fig:setup}). The dipole was placed at a depth $h_d=-20$ Mm, providing the magnetic field strength at the bottom of the numerical domain at $x=-600$ Mm, set to $B=18.9$ G.

 Since the global configuration consists of the superposition of the 2.5D catastrophe and an ordinary dipolar field, the global magnetic field structure includes a null point at a height of $220$ Mm. Located high in the corona, this region is not a subject of this study. Of greater significance is the region between $x = -500$ Mm and $x = -100$ Mm, at heights below 100 Mm, where the magnetic field cancels out. This can be considered representative of the quiet Sun corona. This setup, therefore, provides an opportunity to study the propagation of coronal waves through a highly non-uniformly magnetized corona. Figure \ref{fig:speeds} shows the Alfvén and sound speeds along the horizontal direction at $y=10$ Mm and the vertical direction at $x=0$ Mm. Along the horizontal direction, the Alfvén speed decreases from $1855 \kms$ to $387\kms$ at $x=-100$ Mm (top panel of Fig. \ref{fig:speeds}). Further away from the eruption, the Alfvén speed becomes lower than the sound speed, $c_{s}=150 \kms$. Around $x=-500$ Mm, the Alfvén speed increases again due to the presence of the dipolar magnetic field, reaching $629 \kms$ at $x=-600$ Mm, $y=10$ Mm. The bottom panel of Fig. \ref{fig:speeds} shows a more gradual reduction of the Alfvén speed along the vertical direction as the background quadrupolar magnetic field strength drops.
                                
%boundary conditions and flux rope formation
We used zero-gradient boundary conditions for all variables at the left and right boundaries. This ensured that waves and shock fronts could pass through with minimal reflections, as we used approximate Riemann solver-based discretizations.
  At the bottom, the density and pressure were fixed according to their initial values. The magnetic field was set according to the second-order zero-gradient extrapolation.
 The region of the added dipole field was first subjected to controlled field deformations, generating a flux rope where the prominence was later loaded. This field deformation happened through spatiotemporally prescribed velocities enforced in the local ghost cell regions at the bottom boundary.
The velocities at the bottom ensured the converging and shearing motions ($v_x$ and $v_z$), while antisymmetry was applied for $v_y$. The profile of the converging flow $v_x$ was defined following  Eq. 5 in \citet{Liakh:2020aap} with the center of the profile at $x_c=-600$ Mm, size parameters of the converging region were $W=2\sigma$, $\sigma=21.6$ Mm. The shearing velocity was defined as $v_z=-v_x$. The temporal evolution of these imposed bottom boundary flows was defined by Eqs. 15-17 in \citet{Jenkins:2021aap} but with a smooth profile in time for the deactivation stage. The activation and deactivation times were set to $0$ and $25\mins$, respectively. After the formation of the flux rope, a zero-velocity boundary condition was applied. At the top boundary, the density, pressure, and velocity components were set to match the corresponding values from the last computational cell within the physical domain. The magnetic field components were assigned values from the last cell of the physical domain but with the reverse sign.
 
 %prominence formation
In this study, the prominence was formed by artificially loading plasma into the flux rope dips using a source term in the continuity equation, following an approach adopted in previous studies \citep[e.g.,][]{Liakh:2020aap, Liakh:2021aap, Liakh:2023aap}. The prominence, with a size of $4 \times 4$ Mm and a density of $\rho \approx 10^{-12} \ \mathrm{g\ cm^{-3}}$, was loaded at $x = -600$, $y = 10\ \mathrm{Mm}$ during the time from $25$ to $26.7\mins$. This method was chosen over a more realistic model of prominence formation (e.g., levitation-condensation) due to the time constraint associated with the perturbations from the eruption reaching the prominence region.

                \section{Results}\label{sec:results}
The numerical experiment presented in this paper involved a large coronal region permeated by a complex magnetic structure. Animation 1 associated with Fig. \ref{fig:setup} shows several important processes within this region, including a large-scale eruptive event with an associated extended current sheet underneath, the fragmentation of this current sheet into plasmoids, the propagation of perturbations over large distances, and the interaction of these perturbations with a remote prominence. Studying these regions individually helps to understand the global evolution.
        \subsection{Eruption evolution}\label{sec:eruption}
    \begin{figure*}[!ht]
         \centering
        \includegraphics[width=0.98\textwidth]{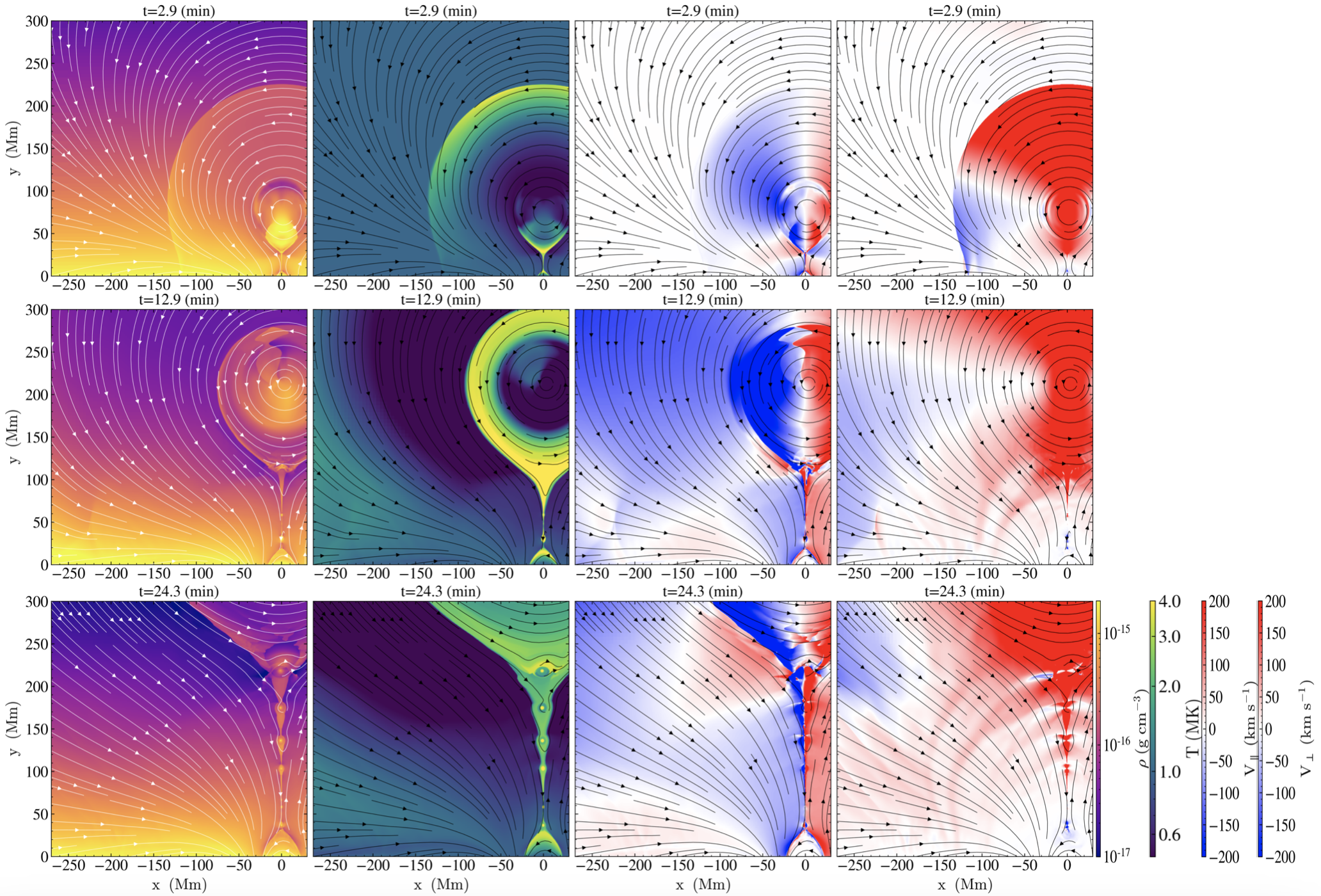}
        \caption{Density, temperature, $v_{\parallel}$, and $v_{\perp}$ distributions during various stages of the eruption: onset (top row), the appearance of the secondary front and fragmentation of the current sheet (middle row), and multiple plasmoid formation in the current sheet (bottom row). Animation 2 shows the temporal evolution up to $57.2\mins$. An animation of this figure is available online.  \label{fig:eruption_evolution}}
\end{figure*}   
\begin{figure*}[!ht]
        \includegraphics[width=0.48\textwidth]{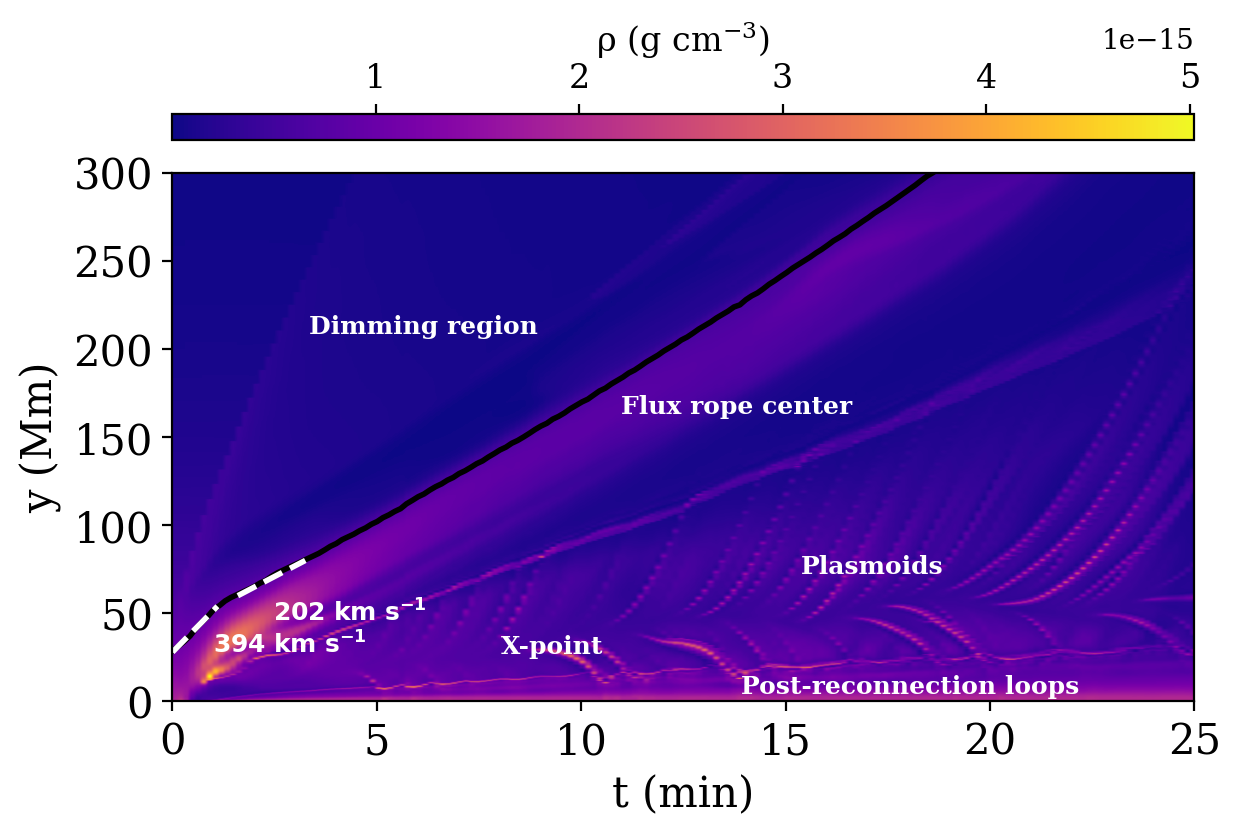}
        \includegraphics[width=0.48\textwidth]{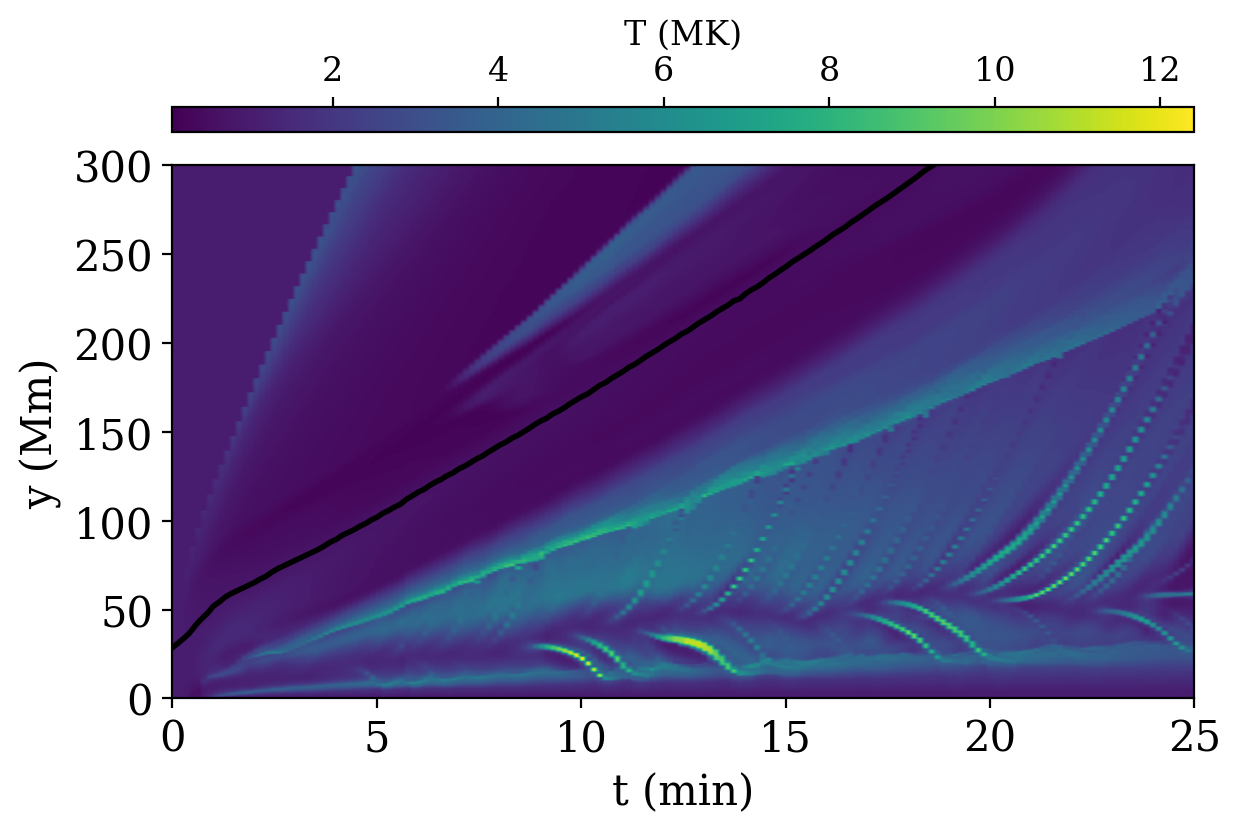}
        \caption{Time-distance diagrams of density (left) and temperature (right) along the vertical cut at $x=0$ Mm. The black line indicates the instantaneous center of the EFR. \label{fig:tdd_density_temperature}}
\end{figure*}
\begin{figure*}[!t]
         \centering
        \includegraphics[width=0.95\textwidth]{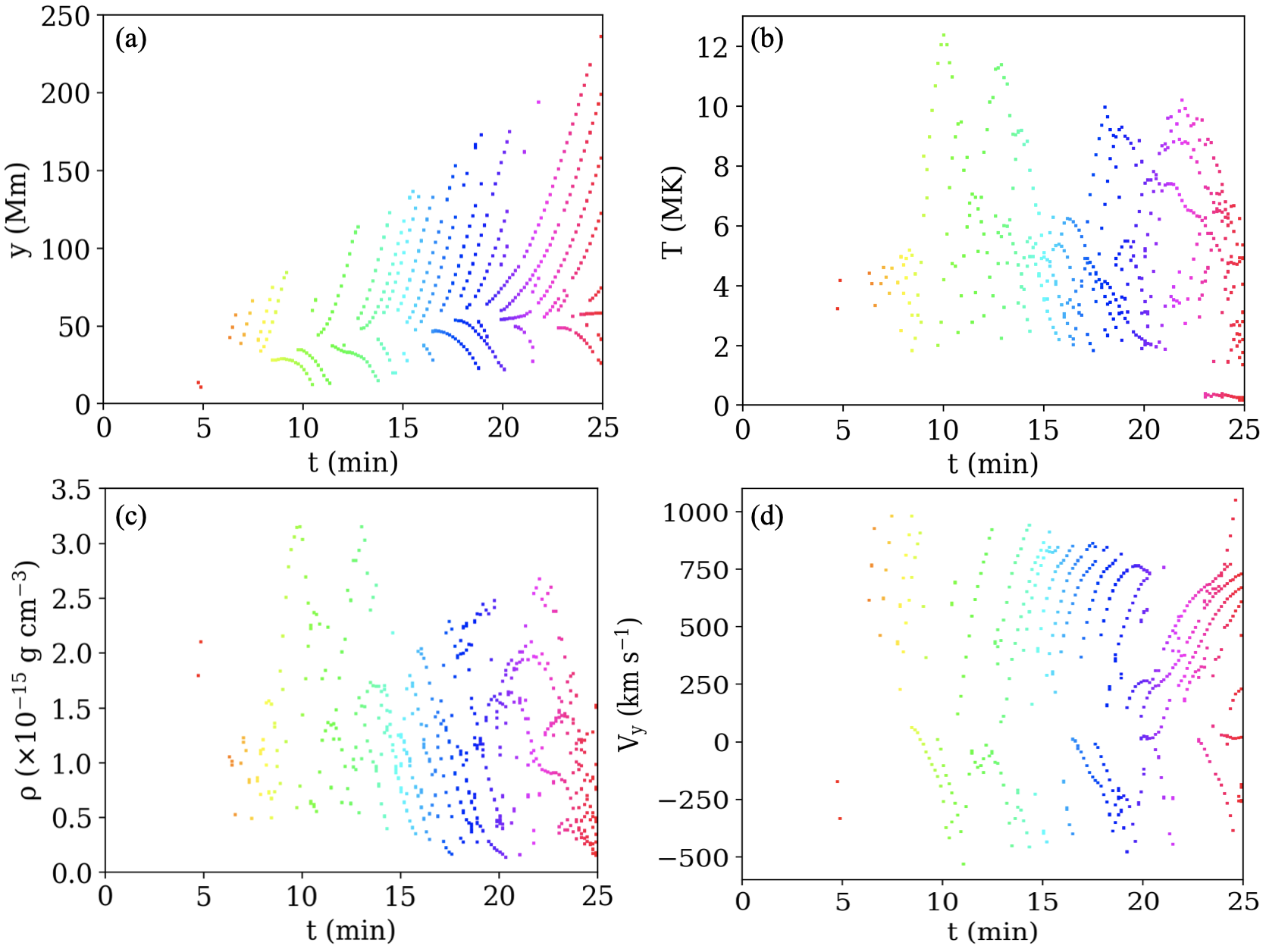}
        \caption{Temporal evolution of plasmoids and their physical properties. Panel (a): Plasmoid trajectories within the current sheet. Panel (b)-(d): Temperature, density, and the vertical velocity $v_y$, corresponding to the instantaneous plasmoid positions. The same color scheme identifies the plasmoids in all the panels.\label{fig:plasmoids_properties}}
\end{figure*}    
The EFR becomes unstable immediately since we chose the catastrophe parameters accordingly. In the top row of Fig. \ref{fig:eruption_evolution}, the EFR has already risen, and the primary front has formed in response to the initial force imbalance. The primary front is not perfectly circular due to variations in physical conditions, such as the density, magnetic field, and, consequently, the Alfvén speed in the vertical and horizontal directions (Fig. \ref{fig:speeds}). We explore this aspect in Sect. \ref{sec:wave}. The top row also shows the magnetic reconnection beginning below the EFR. As a result, the reconnection outflow interacts with the EFR, appearing as a propagating density and temperature perturbation in the lower part of the flux rope. In Animation 2, this perturbation moves along the circular field lines (also visible in the middle row of Fig. \ref{fig:eruption_evolution}).

The middle row of Fig. \ref{fig:eruption_evolution} shows that the EFR has reached a height of $200\Mm$ at $12.9 \mins$. By this time, the primary front has propagated away from the considered region. Two other fronts are still visible in the distributions of the longitudinal and transverse velocities, defined as $v_{\parallel}=(\mathbf{v}\cdot\mathbf{B})/B$ and $v_{\perp}=v_y-v_{\parallel} B_y/B$, around the region from $x=-250$ to $x=-200$ Mm. These two fronts are remnants of the evolution of the primary front. We study these fronts in more detail in Sect. \ref{sec:wave}. In the same row, we observe a region of decreased density and temperature, which separates the EFR from the surrounding corona. Due to its low density, this region appears dark in observations and is referred to as a dimming region \citep[see, e. g.][]{Attrill:2009apj, Dissauer:2018apja, Dissauer:2018apjb, Dissauer:2019apj, Vanninathan:2018apj, Veronig:2019apj, Ronca:2024aap}. 

The current sheet below the EFR elongates and starts fragmenting soon after the initiation of the experiment ($4.7 \mins$). The bottom row of Fig. \ref{fig:eruption_evolution} shows the moment when multiple plasmoids form in the current sheet. These plasmoids show increased density and temperature compared to the surrounding corona. The transverse velocity reveals multiple fronts associated with the motions of the plasmoids. The accompanying Animation 2 shows that plasmoid formation continues until the final stage of the experiment. As the EFR deviates from the strictly vertical direction after $30 \mins$ due to the asymmetry of the magnetic field, the current sheet also inclines to the left.

To understand the temporal evolution of the EFR, primary front, and current sheet along the vertical direction, we obtained time-distance diagrams of the density and temperature at $x=0$ Mm (Fig. \ref{fig:tdd_density_temperature}). Based on the slope of the black line, the initial velocity of the EFR is approximately $394 \kms$. The initial acceleration of the EFR is due to the non-equilibrium caused by an excess Lorentz force. After this stage, the EFR decelerates and rises at a lower speed of $202 \kms$. The density perturbations, caused by the reconnection outflow interacting with the EFR, propagate along the circular field lines on both sides of the EFR. When they meet, they produce a plasma compression evident at the upper part of the EFR at $y=170$ Mm from $t=7.5\mins$.
As the experiment begins, the primary front rapidly moves away from the eruption. A dimming region between the primary front and the upper part of the EFR can be observed in the temperature time-distance diagram and is slightly less pronounced in the density.
As mentioned before, beneath the EFR, a current sheet forms, eventually fragmenting into plasmoids. From the location of the reconnection point in the time-distance diagrams, we note that fewer plasmoids emerge below the reconnection point than above. Those that emerge below the reconnection point merge with the post-reconnection loops underneath, while those above merge with the EFR. 

To define and track the location of the plasmoids over time, we used a criterion based on the horizontal and vertical components of the magnetic field, namely where $B_x$ and $B_y$ simultaneously change sign. The plasmoid trajectories are depicted in panel (a) of Fig. \ref{fig:plasmoids_properties}. The other panels show the temperature, density, and the vertical velocity, $v_y$. The colors represent temporal evolution, allowing us to identify the plasmoids simultaneously present in the current sheet and their corresponding characteristics.
The first plasmoid, marked with red color, forms in the current sheet at $4.7\mins$ and moves downward. The maximum velocity is around $300\kms$, and the temperature exceeds $4$ MK. From time $t=6\mins$, the plasmoid formation becomes more frequent. 
A chain of plasmoids forms between $6$ and $8\mins$ (orange and yellow colors). They move upward, reaching very high velocities $\approx1000\kms$. These plasmoids have an initial temperature above $4$ MK but then slightly cool down. They have a relatively low density around $10^{-15}\ \mathrm{g\ cm^{-3}}$, which also slightly decreases as they move upward.
Between $8$ and $14\mins$, the plasmoids denoted by the light green color are formed both below and above the reconnection point. Those of them moving downward have velocity exceeding $500\kms$ and contain very hot ($12$ MK) and relatively dense ($3.2\times 10^{-15}\ \mathrm{g\ cm^{-3}}$) plasma compared to the surrounding corona.
A second chain of hot and dense plasmoids with characteristics of $10$ MK and $2.5\times 10^{-15}\ \mathrm{g\ cm^{-3}}$ occurs in the time interval $17.5-25\mins$ (from dark-blue to pink colors). 
        
Overall, the plasmoids in our numerical experiment appear to contain and transport upward and downward hot and dense plasma at substantial velocities. Most plasmoids form above the reconnection point and propagate upward following the EFR. These plasmoids accelerate significantly in the current sheet before merging with the flux rope. The downward-moving plasmoids contain denser and hotter plasma in most cases.

     \subsection{Wave propagation}\label{sec:wave}
     \begin{figure*}[!ht]
         \centering
        \includegraphics[width=0.98\textwidth]{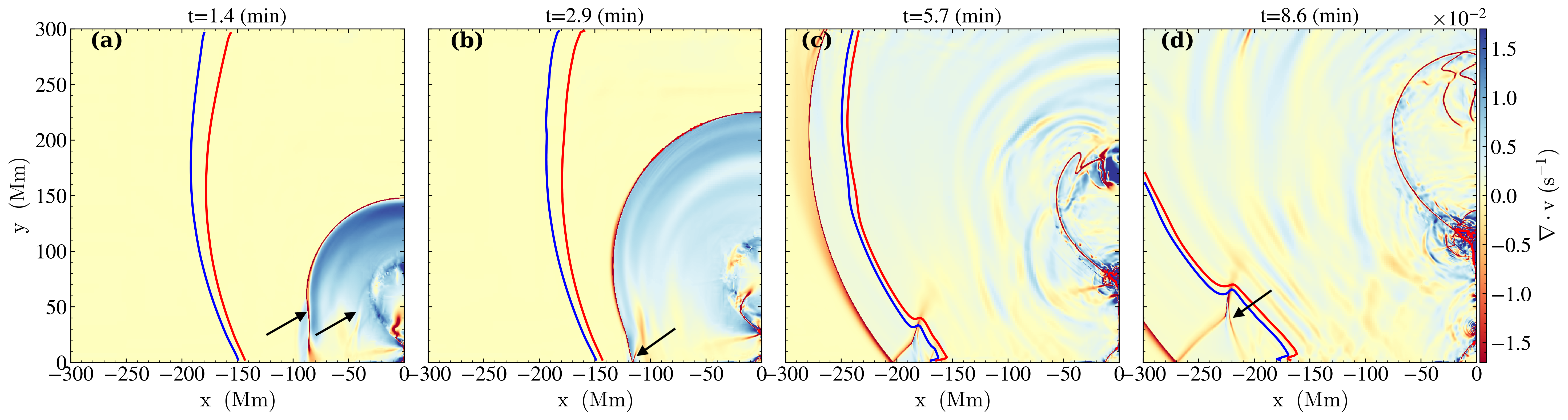}
        \caption{ Temporal evolution of $\nabla\cdot {\mathbf{v}}$. The red and blue contours represent $\beta\approx 1$ and $v_A\approx c_S$ ($\beta\approx 2/\gamma$), respectively. The arrows denote the main fronts detected during the onset of the eruption. \label{fig:divV}}
     \end{figure*}
     \begin{figure}[!t]
        \includegraphics[width=0.45\textwidth]{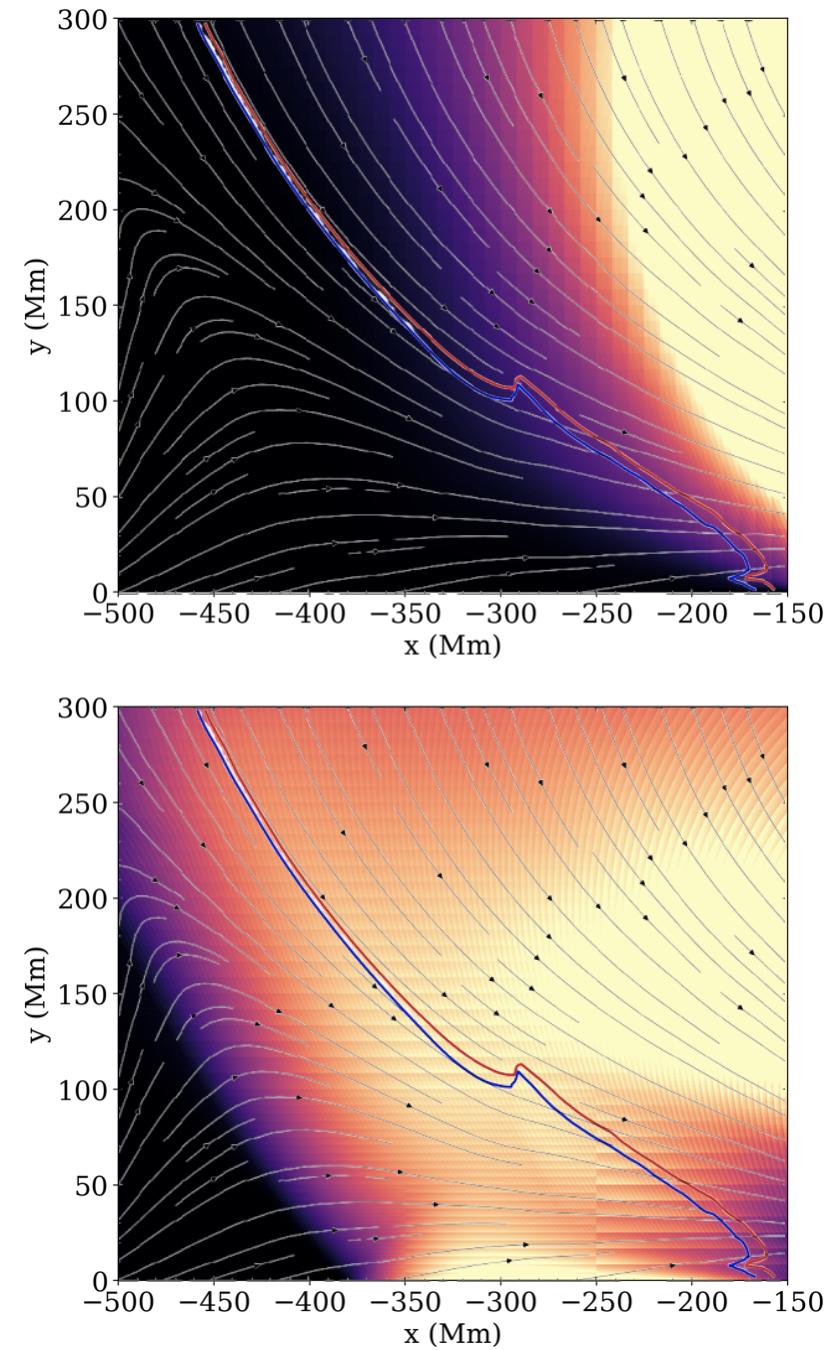}
        \caption{Magnetic (top) and acoustic (bottom) fluxes on the left side of the reconnection site averaged over the first $20\mins$ of the simulation. The gray lines denote the magnetic field lines at $20\mins$. The red and blue contours denote $\beta\approx 1$ and $v_A\approx c_S$ at $t=20\mins$.\label{fig:fluxes}}
     \end{figure}
     \begin{figure*}[!ht]
         \centering
        \includegraphics[width=0.98\textwidth]{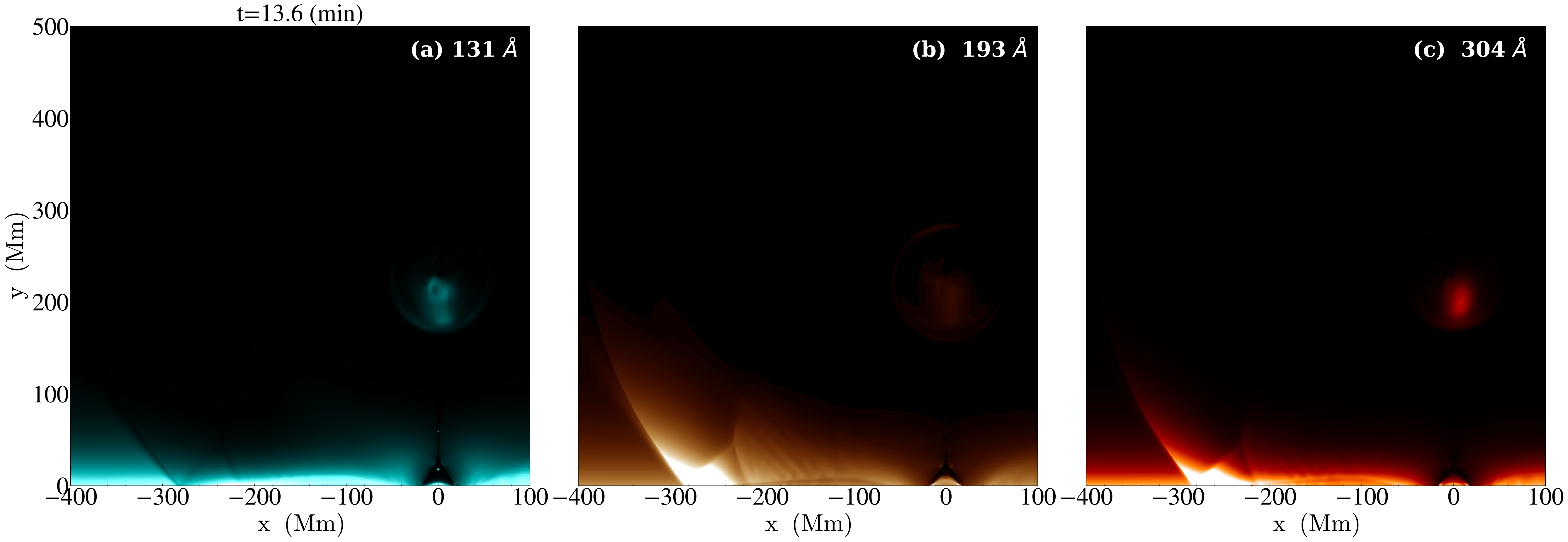}
        \caption{Synthetic SDO/AIA channel images (131, 193, and 304 \AA) of the eruption area at $13.6\mins$. The saturation levels for the fluxes in the 131, 193, and 304 Å channels are defined as follows: $1.5\times 10^{-8}$, $3.0\times 10^{-7}$, $8.0\times 10^{-9}\ \mathrm{DN\ s^{-1}\ pixel^{-1}}$, respectively. We note that these saturation levels are applied to all synthetic images presented in this paper. Animation 3 shows the temporal evolution up to $28.6\mins$. An animation of this figure is available online. \label{fig:aia_eruption}}
     \end{figure*}
     \begin{figure*}[!ht]
        \includegraphics[width=0.49\textwidth]{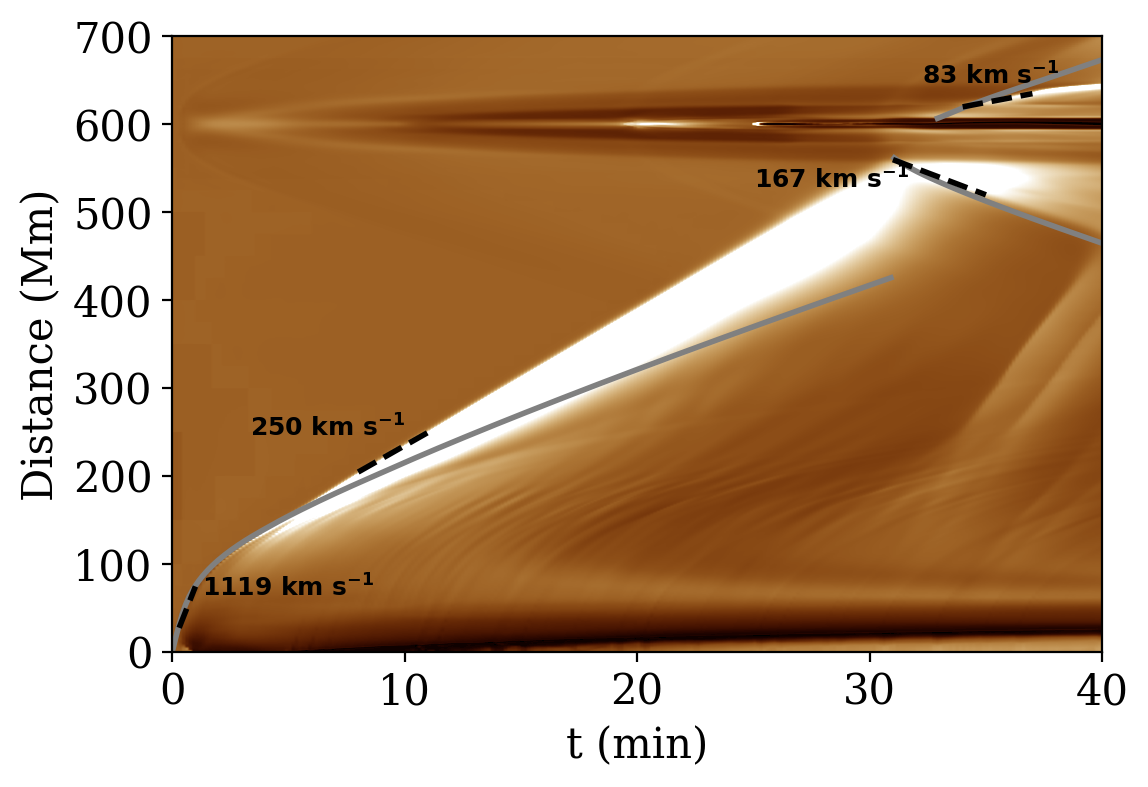}
        \includegraphics[width=0.49\textwidth]{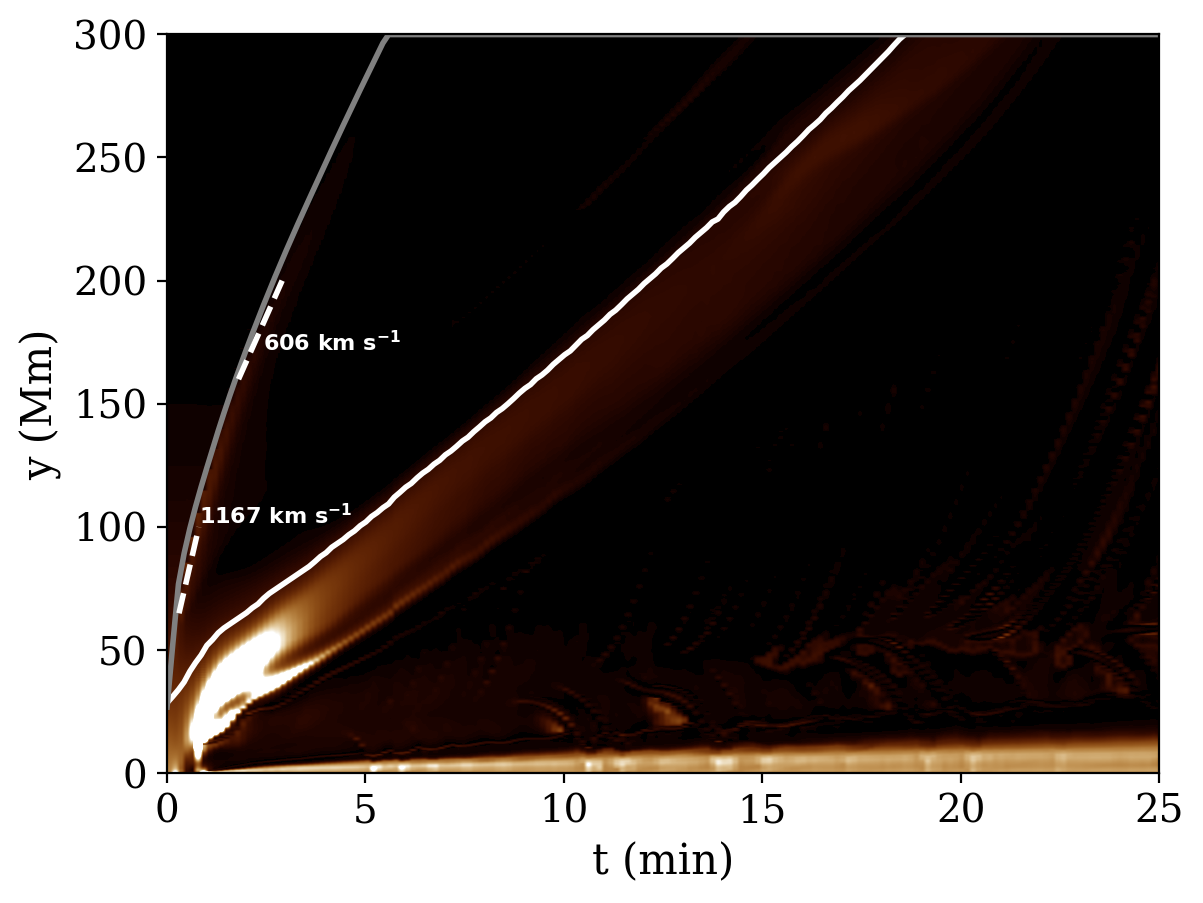}    
        \caption{Time-distance diagrams of 193 \AA\ channel taken along the horizontal cut $y=10$ Mm (left) and the vertical cut at $x=0\Mm$ (right). The vertical axis in the left panel corresponds to the distance from the eruption. The white solid line in the right panel denotes the instantaneous center of the EFR. The gray lines in both panels denote the propagation according to the local phase speed of fast and slow magnetoacoustic waves.\label{fig:tdd_aia193}}
     \end{figure*}
     \begin{figure}[!t]
        \includegraphics[width=0.45\textwidth]{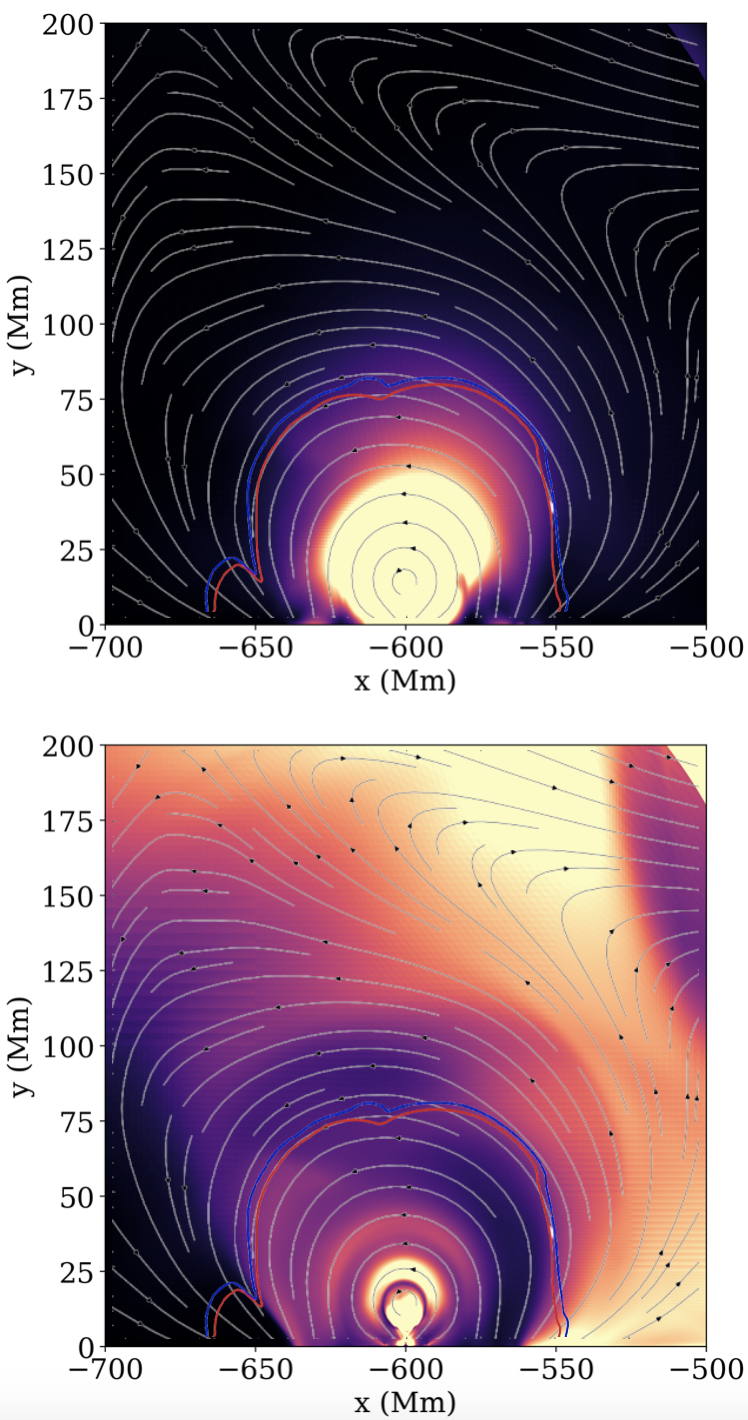}
        \caption{Magnetic (top) and acoustic (bottom) fluxes around the prominence region, averaged over time for the  $20-40\mins$ of the simulation. The gray lines depict the magnetic field lines at $t=40\mins$. The red and blue contours denote $\beta\approx 1$ and $v_A\approx c_S$ at $t=40\mins$. The same saturation levels for the magnetic and acoustic fluxes are applied as in Fig. \ref{fig:fluxes}. \label{fig:fluxes_prominence}}
     \end{figure}
     
The EFR generates various perturbations propagating through the coronal medium permeated by a complex magnetic field. In this section we study the types of perturbations produced and aspects of their evolution. To enhance comparability with observations of such events, we also use synthetic SDO/AIA images.

The velocity divergence, $\nabla\cdot {\mathbf{v}}$, reflects plasma compression, allowing us to detect the fronts produced by the EFR. In Fig. \ref{fig:divV}, the red and blue lines indicate where plasma-$\beta$ is close to unity and where the sound speed is nearly equal to the Alfvén speed. These are critical transitions in the background environment where plasma changes from magnetically dominated to gas dominated. In this experiment, the perturbations are generated in a magnetically dominated region of the EFR, where $\beta << 1$, and then propagate into an area with a weak magnetic field where gas pressure starts to exceed magnetic pressure ($\beta>1$). As the front crosses these equipartition lines, nonlinear effects, such as mode conversion, may occur \citep[see, e.g.,][]{Chen:2016solphys}. This is important for our study, as we aim to understand the potential of coronal waves to probe the magnetic field of the coronal medium. Additionally, a portion of the energy of the primary front can be converted into secondary fronts, thereby weakening the initial perturbation.

Figure \ref{fig:divV}a shows the onset of the eruption, with the EFR center reaching a height of approximately $50$ Mm. At $1.4 \min$, the primary front extends horizontally to about $x=-80$ Mm, and its upper edge reaches $y=150$ Mm.
Figure \ref{fig:divV}a also reveals the presence of another front lagging behind the primary front, with its lateral extent reaching $x=-40$ Mm, although it is less pronounced. Both fronts are indicated by arrows in the panel.
Panel (b) depicts the moment when the primary front impacts the bottom boundary where the zero-velocity condition is applied, producing an echo. 
%As explained by \citet{Wang:2009apj}, this echo results from the interaction between the wavefronts generated by the eruption and the much denser lower atmosphere. 
The echo is evident well before the primary front reaches the equipartition lines.
Panels (c) and (d) show the further evolution of the primary front as it passes the $\beta \approx 1$ and $v_A\approx c_S$ lines, revealing the formation of a secondary front. 
%These panels also reveal multiple fronts surrounding the EFR region generated by plasmoids.
According to \citet{Chen:2016solphys}, a fast-to-slow mode conversion can occur as a front moves from a magnetically dominated plasma to one where gas pressure prevails. To study this further, we analyzed the time-averaged magnetic and acoustic wave fluxes defined by \citet{Bray:1974} as
\begin{eqnarray}
        \mathbf{F_{ac}}=p_1\mathbf{v_1} \, \label{eq:acou_flux}, \\ 
        \mathbf{F_{mag}}=\frac{\mathbf{B_1}\times(\mathbf{v_1}\times\mathbf{B_0})}{4\pi} \, \label{eq:mag_flux},
\end{eqnarray}
where $\mathbf{B_0}$ is the initial magnetic field, and $p_1=p-p_0$ and $\mathbf{B_1}=\mathbf{B}-\mathbf{B_0}$ are perturbations of the pressure and magnetic field with respect to their initial values. Since the initial velocity is zero, we have $\mathbf{v_1} = \mathbf{v}$.
In regions where the Alfvén speed exceeds the sound speed, the magnetic flux can detect a combination of the Alfvén and fast magnetoacoustic modes. The acoustic flux detects the slow magnetoacoustic mode.
We selected an averaging time of $20\mins$, during which the primary front propagates through the region of interest shown in Fig.~\ref{fig:fluxes}, and the secondary front has already formed. Averaging the fluxes over multiple wave periods (approximately $5\mins$ each) allows us to capture the energy transfer dynamics and cumulative effects.

As shown in the top panel of Fig. \ref{fig:fluxes}, the magnetic flux is predominantly concentrated along the path of the flux rope and the upward propagation of the main front. In this same region, the acoustic flux is mainly distributed around a height of $150$ Mm. This distribution may be due to the rising EFR, which contains plasma with significantly higher pressure than the surrounding environment. Notably, the bottom panel reveals another enhancement in the acoustic flux extending along the current location of the equipartition lines (indicated by the red and blue lines). All described above makes us conclude the secondary front observed in Fig. \ref{fig:eruption_evolution} is the slow magnetoacoustic wave formed when the primary front crosses the equipartition lines.  The bottom panel of Fig.~\ref{fig:fluxes} shows the lateral propagation of the fast magnetoacoustic wave beyond the positions of the red and blue lines. In the regions where the sound speed exceeds the Alfvén speed, the acoustic flux indicates the propagation of the fast magnetoacoustic wave.

To improve comparability with real observations, we generated synthetic images using the \texttt{MPI-AMRVAC} code as described by \citet{Xia:2014apjl} in three SDO/AIA channels to cover different temperature ranges: 304 Å (primarily capturing temperatures around $0.08$ MK), 193 \AA\ ($1.5$ MK), 131 \AA\ (primary temperature response: $10$ MK and secondary: $0.4$ MK). 
%131 A channel
The eruption is seen in the 131 \AA\ channel mostly due to the flows and compressions produced by the interaction of the reconnection outflow with the EFR. From Fig. \ref{fig:eruption_evolution}, we know that the temperature of this flowing plasma inside the EFR varies in the range $0.6-4.0$ MK. Therefore, the plasma inside the EFR becomes visible due to the secondary peak of the 131 \AA\ channel. In the 131 \AA\ channel, we can observe post-reconnection loops and plasmoids with high temperatures of $10$ MK. 

%193 A channel
The primary front appears bright in the 193 \AA\ channel in Fig. \ref{fig:aia_eruption}, indicating a plasma temperature of around $1.5$ MK. A secondary front is also identifiable at around $x=-200$ Mm, corresponding to the front in Figs. \ref{fig:eruption_evolution}(middle row), \ref{fig:divV}d and \ref{fig:fluxes}(bottom). Additionally, multiple fainter fronts are produced by plasmoids in the region between $x=-200$ and $-100$ Mm. In the corresponding Animation 3, these fronts become less visible and are difficult to identify at distances around $400$ Mm from the reconnection site. They, therefore, cannot affect the distant prominence. Animation 3 also shows the formation of the dimming and bright compression regions surrounding the EFR, most clearly seen at $t=4.3\mins$.

%304 A channel
Finally, the 304 \AA\ channel corresponds to lower temperatures around $0.08$ MK. From Fig. \ref{fig:aia_eruption} and the corresponding Animation 3, this channel also shows details such as the flows and compressions inside the EFR, post-reconnection loops, and the primary and secondary fronts. 

To analyze primary front propagation in the horizontal and vertical directions, we obtained time-distance diagrams in the 193 \AA\ channel:  a horizontal diagram at $y=10$ Mm, corresponding to the height of the prominence center set by the mass loading process explained in Sect. \ref{sec:numerical} and a vertical diagram at $x=0$. 
%
%Plasmoids are also visible in this time-distance diagram, although they are more prominent in the 131 \AA\ time-distance diagram as their temperatures exceed $10$ MK (not shown here). 
%
The left panel of Fig. \ref{fig:tdd_aia193} illustrates the propagation of the primary front in the horizontal direction. The initial speed of the front exceeds $1000\kms$, but due to the rapid decrease in the magnetic field, the front quickly decelerates to $250\kms$. The gray line shows the slope given by the local phase speed of the fast magnetoacoustic wave $v_{ph}=\sqrt{v_{A}^2+c_{s}^2}$. Initially, the slopes coincide up to time $5\mins$. However, at a distance of $150$ Mm, the primary front deviates from the gray line. This analysis suggests that the primary front initially propagates laterally as an ordinary fast magnetoacoustic wave and later becomes a shock wave. 
%Behind the shock front is the brightening linked to plasma piling. 
%
%A secondary front, discussed previously, forms at approximately $x=150-200$ Mm and appears faint. 
%
The prominence region is located at a distance of $600$ Mm. The primary front reaches the prominence at approximately $30\mins$. As a result of the front interacting with the prominence region, a reflected front with a speed of $167\kms$ is visible. Again, we compare the propagation of this reflected front to the fast magnetoacoustic mode and find agreement. Additionally, a transmitted front forms with a slope velocity of $v=83\kms$, significantly slower than the primary front. We compare this transmitted front with the slow magnetoacoustic wave propagating at local sound speed in the region where $\beta<<1$. Initially, the slopes coincide, but later, this transmitted front slows down, appearing almost as a horizontal line and indicating a stationary front. 

To this end, we also analyzed the magnetic and acoustic fluxes, similar to Fig. \ref{fig:fluxes}, but centered around the prominence region and time-averaged throughout $20-40\mins$. In order to compute the fluxes using Eqs. \ref{eq:acou_flux}, \ref{eq:mag_flux}, the initial values of the pressure, velocity, and magnetic field were assumed to be at  $20\mins$. The top panel of Fig. \ref{fig:fluxes_prominence} shows a bright region, indicating an increase in magnetic flux, which indicates propagation of fast magnetoacoustic and Alfvén waves in the magnetically dominated region at the area of the prominence magnetic field. The bottom panel shows a bright front localized along the loops overlying the flux rope prominence region around $x=-640$ Mm. This front was previously discussed regarding the left panel of Fig. \ref{fig:tdd_aia193}. In the magnetically dominated region, the acoustic flux indicates the propagation of the slow magnetoacoustic wave. Everything described above leads us to conclude that this front is formed by the fast-to-slow mode conversion. It propagates along the loops, finally creating a stationary front.

The right panel of Fig. \ref{fig:tdd_aia193} shows the vertical cut, highlighting similar features discussed in Fig. \ref{fig:tdd_density_temperature}. In this time-distance diagram, the primary front produced by the eruption is identifiable. The estimation of the slope velocities shows that the primary front has an initial speed exceeding $1000\kms$ but decelerates to $606\kms$. The gray line indicates the fast magnetoacoustic wave propagation at the local phase speed. The two slopes coincide, indicating that the primary front propagates upward as an ordinary fast magnetoacoustic wave.

%Another mode conversion happens due to the passing of the fast magnetoacoustic wave from the magnetically dominated region to the gas pressure-dominated region, i.e., the opposite of the case discussed in Fig. \ref{fig:fluxes}.

    \subsection{Prominence evolution}\label{sec:prominence}
   %        \begin{figure*}[!ht]
   %    \includegraphics[width=0.98\textwidth]{density_temperature_velocity_prominence_interaction_plot.png}
    %   \caption{Density, temperature, $v_{\parallel}$, and $v_{\perp}$ distributions during various stages of the experiment, focusing especially on the deformed dipolar region that hosts the prominence: prominence mass loading (first row), the wave interaction with the prominence (second row), prominence oscillations and mass accretion (third and fourth rows). The black arrow denotes the primary front. Animation 4 shows the temporal evolution up to $57.2\mins$. (An animation of this figure is available online.) \label{fig:prominence_evolution}}
   %\end{figure*}
    \begin{figure*}[!ht]
        \centering
        \includegraphics[width=0.9\textwidth]{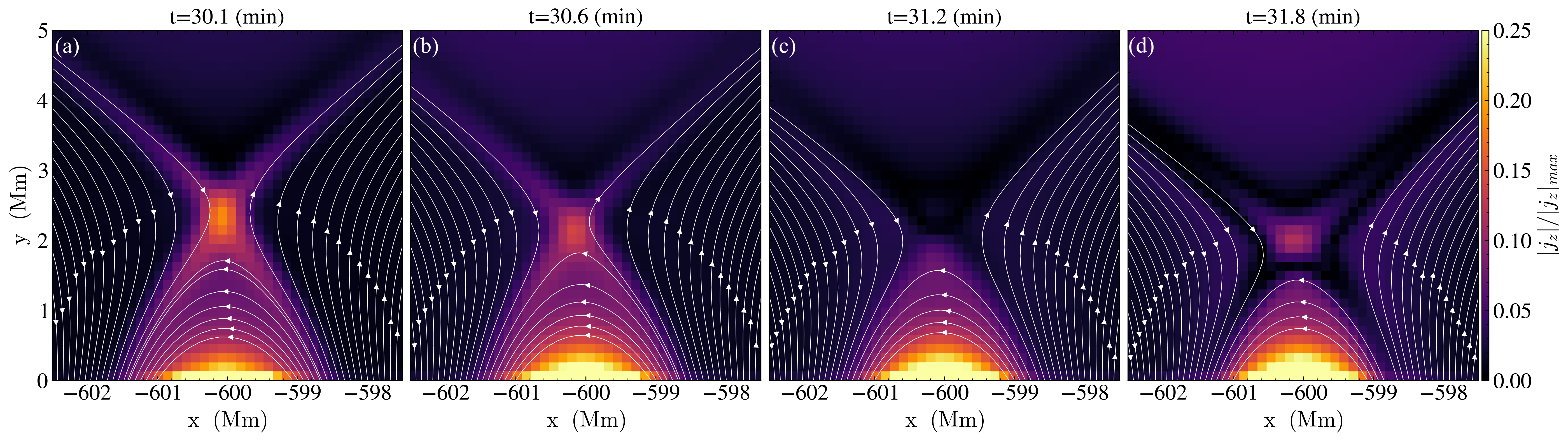}
        \caption{Zoomed-in details of the wave front interaction with the remote flux rope, focused on the X-point below the prominence hosting flux rope. The temporal evolution of the current density, $|j_{z}|$, is shown normalized to the instantaneous maximum value in the region, $|j_{z}|_{max}$, along with the magnetic field lines at the bottom of the flux rope after its interaction with the primary front. The white lines in each panel denote the instantaneous position of the same magnetic field lines. Animation 5 shows the temporal evolution in the time interval  $28.6-50.1\mins$. An animation of this figure is available. \label{fig:prominence_evolution_current}}
    \end{figure*}
    \begin{figure}[!t]
        \includegraphics[width=0.45\textwidth]{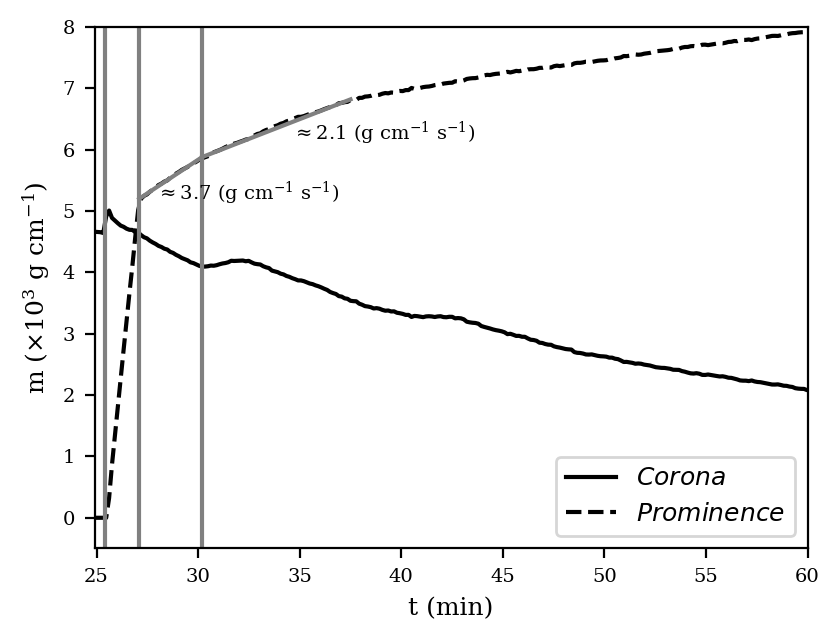}
        \caption{Temporal evolution of the total mass of the coronal and prominence plasma using the threshold, $\rho=10^{-14}\ \mathrm{g\ cm^{-3}}$ in the dynamically evolving and tracked flux rope region.  The vertical gray lines correspond to the activation and deactivation of mass loading and the arrival of the primary front. \label{fig:mass}}
    \end{figure}
    \begin{figure*}[!ht]
         \centering
        \includegraphics[width=0.95\textwidth]{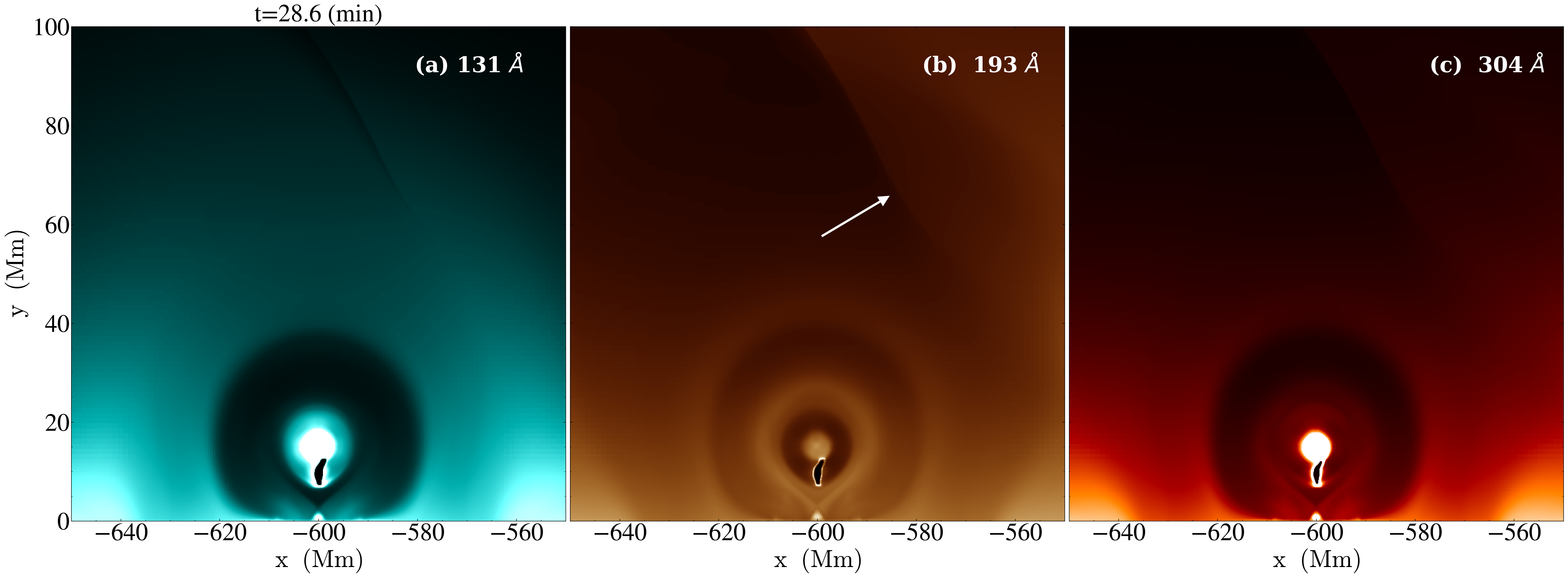}
        \caption{Synthetic images of the prominence region in SDO/AIA channels 131, 171, and 193 \AA, shown at $28.6\mins$. The white arrow denotes the primary front. Animation 6 shows the prominence formation, the passing of the primary front, and the induced prominence dynamics up to $57.2\mins$. An animation of this figure is available online.\label{fig:prominence_evolution_aia}}
    \end{figure*}
    \begin{figure}[!h]
        \includegraphics[width=0.45\textwidth]{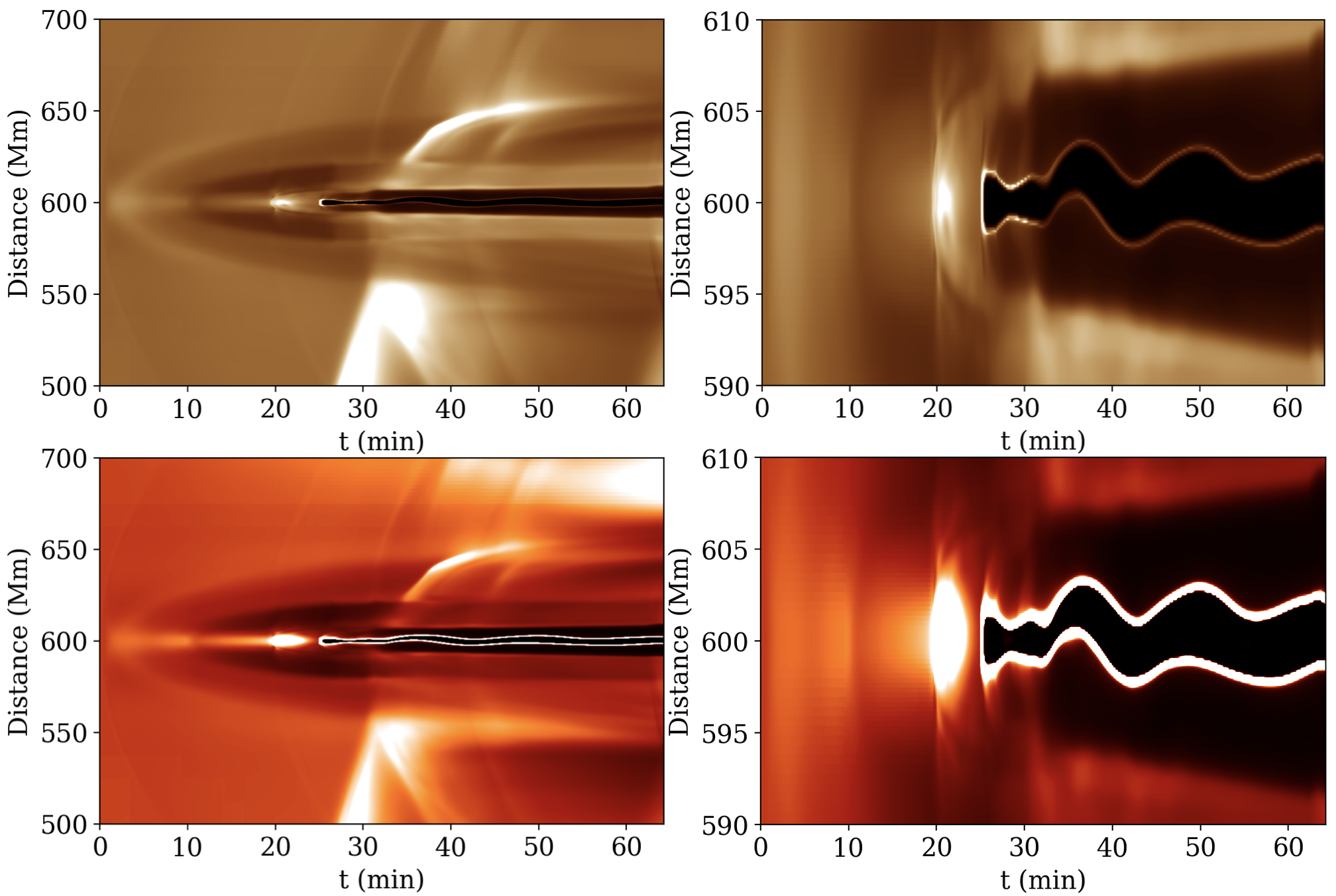}
        \caption{Time-distance diagrams of the 193 \AA\ (top) and 304 \AA\ (bottom) SDO/AIA channels taken along the horizontal cut at $y=10\Mm$. The right panels show a zoomed-in view around the prominence region. The vertical axes denote the distance from the eruption. \label{fig:time_distance_diagram_prominence}}
    \end{figure}
    \begin{figure}[!h]
        \includegraphics[width=0.45\textwidth]{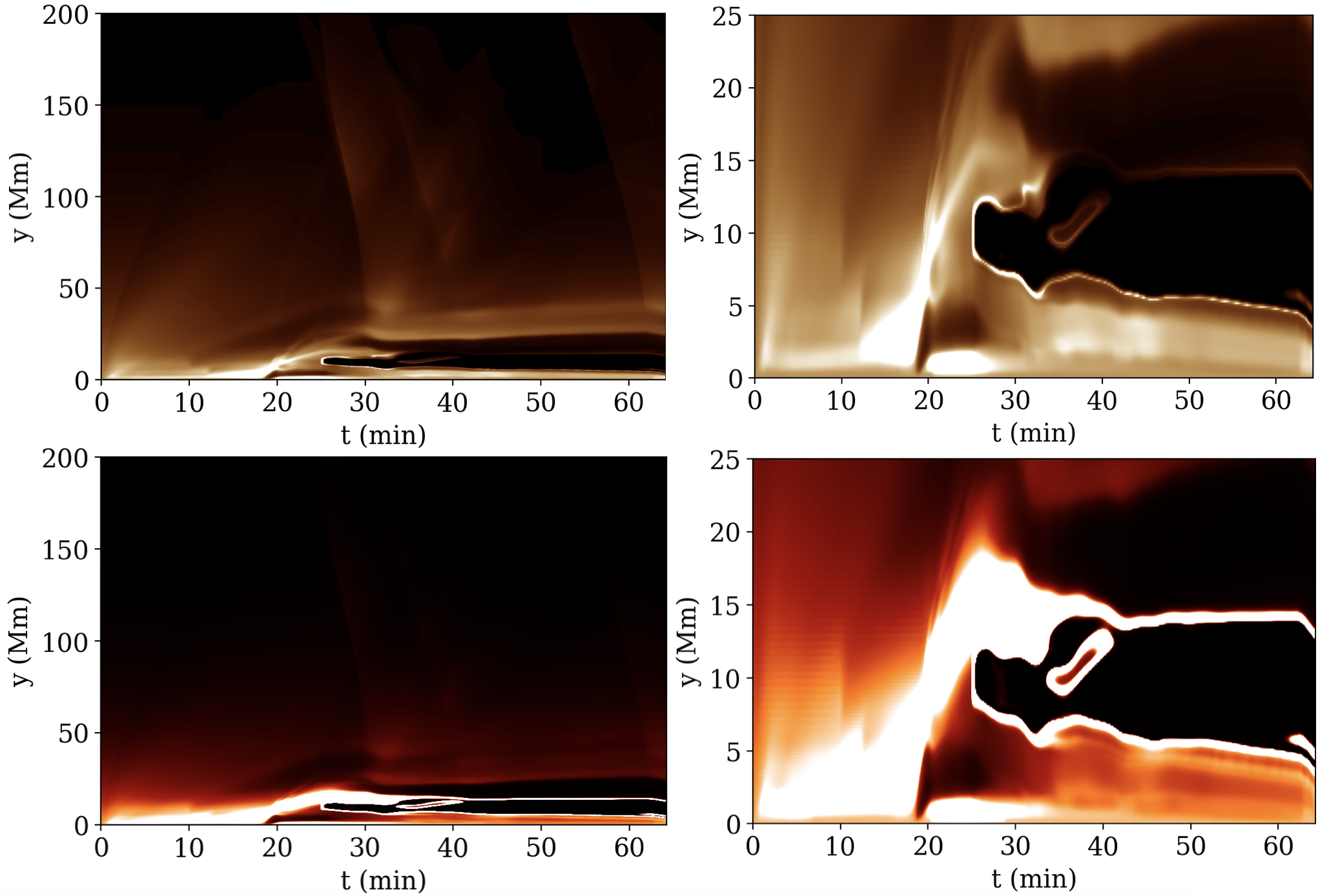}
        \caption{Time-distance diagrams of the 193 \AA\ (top) and 304 \AA\ (bottom) SDO/AIA channels taken along the vertical cut at $x=-600\Mm$. The right panels show a zoomed-in view around the prominence region. \label{fig:time_distance_diagram_prominence_ver}}
    \end{figure}
    \begin{figure}[!t]
        \includegraphics[width=0.45\textwidth]{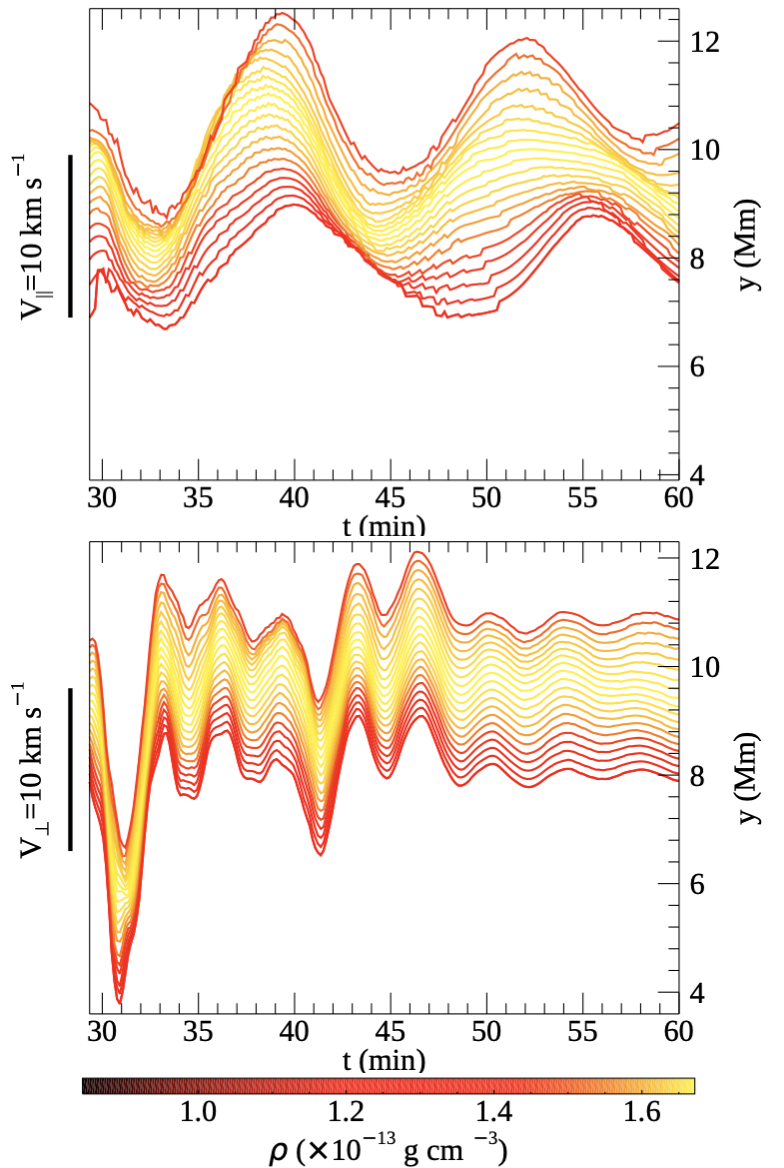}
        \caption{Temporal evolution of $v_{\parallel}$ (top) and $v_\perp$ (bottom) obtained by tracking 20 fluid elements. The right vertical axis shows the initial height of the corresponding fluid element. The color bar corresponds to the initial density at the positions of the fluid elements.\label{fig:line_analysis}}
    \end{figure}
    
We examined the evolution of the eruption, the resulting perturbations, and their propagation throughout the numerical domain. Another aim of this study is to understand how these eruption-driven perturbations interact with a distant prominence. In this section we study the flux rope formation, the loading and evolution of the prominence mass within the flux rope, and the dynamic response of the prominence to the primary front. For this purpose, we analyze key parameters such as temperature, density, and longitudinal and transverse velocities. Additionally, we study the appearance of the flux rope and prominence in synthetic images in the 131, 193, and 304 Å channels.

Figure \ref{fig:prominence_evolution} and the corresponding Animation 4 depict the evolution around $x=-600$ Mm, from the formation of the flux rope and prominence to their late-stage evolution after the passing of the primary front.
The first row of Fig. \ref{fig:prominence_evolution} shows the moment when the flux rope is fully formed. At the center of the flux rope, there is a region of slightly higher density and lower temperature compared to its surroundings. This plasma originates from the lower corona and is lifted during the flux rope formation process. The prominence plasma, artificially loaded into the flux rope, appears as a dense and cold plasma block. The longitudinal velocity indicates that the prominence plasma begins to be compressed toward the center of the magnetic dips.
The second row of Fig. \ref{fig:prominence_evolution} shows that the prominence has a narrower shape and higher density due to compression. Additionally, a `tail' is formed at the top part of the prominence. This tail extends along the circular field lines toward the flux rope center. The density distribution shows that lifted plasma remains mainly close to the flux rope center. The longitudinal velocity suggests that plasma continues flowing towards the main prominence body along the magnetic field. 
At this stage, the primary front from the eruption reaches the prominence region, as shown by the transverse velocity and indicated by the arrow in the last panel of the second row in Fig.~\ref{fig:prominence_evolution}.
The interaction of the wave with the prominence is evident in Animation 4, associated with Fig. \ref{fig:prominence_evolution} and Animation 1, associated with Fig. \ref{fig:setup}. From Animation 1, the bottom part of the primary front becomes oblique before the interaction. This happens because the density decreases more rapidly than the magnetic field along the vertical direction between $x = -500$ and $-100$  Mm. As a result, the Alfvén speed increases with height, leading to the inclination of the front.

As shown in Animation 4, the flux rope and prominence are displaced to the left and down during the propagation of the primary front. As a result, the prominence starts to move around the center of the magnetic dips. Additionally, when the flux rope is pushed down, we can see a gradual disappearance of the post-reconnection loops below the flux rope. Figure \ref{fig:prominence_evolution_current} shows the details of the evolution of the current density and the magnetic field in the null point region. In order to follow the evolution of the same magnetic field lines each time moment, we started the integration at the bottom, where the zero-velocity condition is applied. Panel (a) shows the small vertical current sheet below the prominence just before the arrival of the primary front. When the front pushes the flux rope down, the current density of the vertical sheet starts to decrease (Fig. \ref{fig:prominence_evolution_current}b). Panel (c) shows the moment when the vertical current sheet entirely disappears. Finally, in panel (d), the formation of the horizontal current sheet is seen as the current density increases. The magnetic field lines of the post-reconnection loops reconnect at the horizontal current sheet and form the overlying magnetic field lines, which move gradually away from the reconnection region. The corresponding Animation 5 shows more details of the evolution of the null point, particularly that all the post-reconnection loops denoted in Fig. \ref{fig:prominence_evolution_current}a are reformed by reconnecting with the field lines of the flux rope.
\citet{McLaughlin:2004aap} highlighted the critical role of null points in the dissipation of fast magnetoacoustic waves, attributing this to increased Ohmic heating resulting from enhanced currents. Subsequent studies revealed that nonlinear waves can deform null points, causing them to collapse into current sheets. This process leads to a sequence of horizontal and vertical current sheet formations, the process commonly referred to as oscillatory reconnection \citep{Mclaughlin:2009aap, Mclaughlin:2012aap}. Oscillatory reconnection can be interpreted as damped harmonic oscillations. The results of our experiment resemble this behavior, as we observe rapidly damped oscillatory reconnection.

The third and fourth rows of Fig. \ref{fig:prominence_evolution} show the prominence evolution towards the end of the numerical experiment. The fourth row shows that the prominence motions are almost entirely damped. The lifted plasma contained in the center of the flux rope is seen in a tiny region in the third row and entirely merges with the main prominence body in the fourth row. Another tail forms at the prominence bottom, and the longitudinal velocity shows that there is still plasma flowing towards the main prominence body from the edges of the flux rope.

Figure \ref{fig:mass} quantifies the evolution of the total mass inside the flux rope after it is fully formed ($25\mins$). We defined the instantaneous center of the flux rope as the maximum of the current density and identified the flux rope region with a local ellipsoid. Thus, at each time step, the center of this local ellipsoid dynamically evolved along the vertical direction at $x=-600$ Mm. The axes of the ellipsoid, which defined the shape of the magnetic field lines, were fixed to be $24.1$ and $14.3$ Mm in the vertical and horizontal directions, respectively. Then, we integrated the plasma density contained inside the flux rope, distinguishing between the coronal and the prominence plasma using a density threshold of $\rho_{crit}=10^{-14} \ \mathrm{g\ cm^{-3}}$. Figure \ref{fig:mass} shows that at $25\mins$, the total mass of the coronal plasma is $4.6\times 10^{3} \ \mathrm{g\ cm^{-1}}$. The prominence plasma is gradually loaded during $1.67\mins$, marked by the first two vertical gray lines. As this loaded plasma occupies the region initially filled with the coronal plasma, the total amount of the coronal plasma decreases. After the prominence is fully loaded, its total mass slowly increases, reaching $8\times 10^{3} \ \mathrm{g\ cm^{-1}}$ at the final stage of the experiment. Assuming that the length of the prominence in the third direction can be as long as $100$ Mm, the total prominence mass is $\approx 10^{14}\ \mathrm{g}$, similar to that obtained in the recent 3D prominence simulation \citep{Donne:2024apj}. Simultaneously, the total mass of the coronal plasma decreases until it reaches $2\times 10^{3} \ \mathrm{g\ cm^{-1}}$. The variation of the total masses reflects the accretion of the coronal plasma contained inside the flux rope towards the artificially loaded prominence. From this figure, we can see that the accretion rate is $3.7\ \mathrm{g\ cm^{-1}\ s^{-1}}$ and  $2.1\ \mathrm{g\ cm^{-1}\ s^{-1}}$ before and after the passing of the primary front. From this analysis, we do not see a strong influence of the primary front on the accretion rate inside the flux rope.

 Figure \ref{fig:prominence_evolution_aia} and the associated Animation 6 show the brightening in all channels within the flux rope core, where the lifted plasma is localized. 
The flux rope formation leads to the depletion of the overlying loops. Therefore, the magnetic loops surrounding the flux rope appear dark in all channels.
The cold prominence plasma, loaded inside the flux rope, also appears dark in these channels. The prominence edges appear bright, reflecting the relatively hot prominence-corona transition region (PCTR). The PCTR consists of plasma with temperatures spanning the range between prominence and coronal values, which explains its brightening in the 131, 193, and 304 Å channels. The temperatures of the PCTR correspond to the peak of the chosen radiative cooling curve, $\Lambda(T)$ \citep[see Fig. 1 in][]{Hermans2021}. Given the relatively high density in the PCTR compared to the corona, this region is influenced by the significant radiative cooling, governed by the $\rho^2\Lambda(T)$ term. This cooling inevitably leads to a decrease in gas pressure within the PCTR. To restore gas pressure balance, dense plasma contained within the flux rope begins to accrete onto the PCTR. Therefore, after the initial brightening of the lifted plasma inside the flux rope, visible in all channels, this brightening gradually fades, reflecting the plasma accretion onto the main prominence body. Depleting the flux rope leads to the formation of a dark coronal cavity in all the considered SDO/AIA channels. The cavity grows as more plasma accretes on the main prominence body, and this process continues until the end of the numerical experiment.

Between 28 and 50 minutes, the passing of the EUV front is seen in the 193 Å channel (the white arrow in Fig. \ref{fig:prominence_evolution_aia}b). As the primary front passes, it interacts with the dark cavity and overlying loops. Notably, the front leaves behind a bright transmitted front that stops in the dark overlying loops. 
%As previously discussed regarding the left panel of Fig. \ref{fig:tdd_aia193}, this front forms due to fast-to-slow mode conversion.
%
After the front passing, the prominence oscillates, a motion primarily visible due to the brightness of the PCTR. After a few minutes, these oscillations are damped. 
Additionally, the EUV front triggers magnetic reconnection, which was discussed earlier. This causes the bright post-reconnection loops to gradually disappear across all channels.

To investigate the interaction between the EUV front and the prominence, we generated time-distance diagrams for the 193 and 304 Å channels along the horizontal direction at $y=10$ Mm (Fig. \ref{fig:time_distance_diagram_prominence}) and the vertical direction at $x=-600$ Mm (Fig. \ref{fig:time_distance_diagram_prominence_ver}). Figure \ref{fig:time_distance_diagram_prominence} captures the flux rope formation during $0-25$ minutes, showing the top part of the flux rope crossing $y=10$ Mm, followed by the bright central region of the compressed and lifted plasma, and finally, the bottom part of the flux rope (i.e., the magnetic dips). Shortly after this, the mass-loading process begins and lasts about $1.67$ minutes. The loaded plasma appears dark in both channels except for the PCTR. After the prominence is loaded, it undergoes a cycle of compression and rarefaction, which is disrupted by the arrival of the EUV front at around $30\mins$ (see left panels in Fig.~\ref{fig:time_distance_diagram_prominence}). Upon impact of the primary front, the prominence is displaced at around $3-4$ Mm, and the oscillations are initiated. As plasma accretes onto the main prominence body, the region between $592$ and $608$ Mm becomes dark in both channels. 

Figure \ref{fig:time_distance_diagram_prominence_ver} shows the evolution along the vertical cut. In the left panels, the primary EUV front passes the prominence region at $30\mins$. The right panels show zoomed views capturing the formation and rise of the flux rope. The brightening occurs due to plasma collected during flux rope formation. Later, this plasma is lifted by the flux rope, reaching heights of $10-20$ Mm by $t=25\mins$. The prominence mass is loaded at $25\mins$ and leads to a slight descent of the flux rope. The primary front arrives at $30\mins$, pushing the flux rope downward and initiating motions in the vertical direction. These motions are damped entirely by $45\mins$. The right panels also show bright plasma accreting onto the prominence. This leads to the formation of the dark cavity above $y=12-15$ Mm. The vertical size of the prominence significantly increases as a result of the mass accretion.

From the analysis described above, we conclude that the primary front from the eruption induces prominence plasma motion by displacing it to the left and simultaneously pushing it downward. To analyze local plasma motion, we could examine advection along the magnetic field lines, similar to previous studies \citep[see, e.g.,][]{Liakh:2020aap}. However, this method is not applicable when the magnetic field evolves significantly, as in this study. Therefore, we analyzed the longitudinal and transverse velocities interpolated along the trajectories of fluid elements, as performed in \citet{Liakh:2023aap}. We selected 20 fluid elements at $x=-600$ Mm and $y=7.8-10.8$ Mm at $t=29.6$ minutes, just after the prominence loading but before the perturbing front reaches this region. In our analysis, the longitudinal velocity reflects prominence oscillations around the centers of the magnetic dips. The transverse velocity, on the other hand, represents oscillations in the vertical direction. The results of this analysis are shown in Fig. \ref{fig:line_analysis}. The top panel shows the evolution of the longitudinal velocity. When the oscillations are established, their amplitudes vary with height from $4.2$ to $14.0\kms$. We estimated the periods of these oscillations using the Lomb-Scargle periodogram. The period of these oscillations decreases with height, ranging from $15.1$ to $12.4\mins$. %Consequently, the top and bottom motions become unsynchronized by $t=50\mins$. 
We computed the time-averaged radii of the curvature of the magnetic field lines for each fluid element, $R_{c}=3.5-4.5$ Mm, in order to obtain the pendulum period, $P_{pendulum}=2\pi\sqrt{R_{c}/g}$ as defined by \citet{Luna:2012apjl}, where $g$ is the solar gravitational acceleration. The pendulum period also decreases with height from $13.4$ to $11.9\mins$. 

We also estimated the damping time using the fitting of the longitudinal velocity with a damped sinusoid function. The damping time slightly decreases with height, ranging from $20$ to $40$ minutes. In our experiment, the plasma accretion on the main prominence body is observed. Therefore, we estimated the damping time of the longitudinal oscillations caused by mass accretion using $\tau_D=1.72\ m_0/\dot{m}=78.5\mins$, where $m_0$ is the total prominence mass before perturbation, and $\dot{m}$ is the mass accretion rate \citep[see Eq. 64 in][]{Ruderman:2016aap}. This damping time exceeds the one we obtained in our experiment, suggesting that the damping mechanism of the longitudinal oscillations cannot be explained only by plasma accretion.

The transverse velocity shows a more significant response to the perturbation caused by the primary front. When the flux rope is pushed downward, the initial transverse velocity is around $12.6-14.2\kms$. These oscillations show more global character with a constant period of $3.6-3.7\mins$ at all heights. To interpret the period, we used the formula from \citet{Hyder:1966zap} for transverse oscillations in the case of a simple harmonic oscillator, where magnetic tension acts as the main restoring force: $P = \frac{2\pi \langle h \rangle}{\langle B \rangle} \sqrt{\pi \langle \rho_p \rangle } = 3.3\mins$, where $\langle h\rangle = 8$ Mm, $\langle B \rangle = 13.4$ G, and $\langle \rho_{max} \rangle = 9 \times 10^{-14}\ \mathrm{g\ cm^{-3}}$ are the prominence height, magnetic field strength, and maximum density, respectively, averaged over the time interval of $30 - 60\mins$. These oscillations are almost entirely damped by $40\mins$, and the damping time is estimated to be around $4\mins$.

\section{Discussion}\label{sec:discussion}

In this paper, we performed a 2.5D numerical experiment using the \texttt{MPI-AMRVAC 3.1} code to investigate the interaction between eruptive events and remote prominences. The model incorporated a 2.5D catastrophe magnetic field to generate the eruption and a dipole field to form the flux rope with an embedded prominence. This magnetic configuration allowed us to study the dynamics of an energetic eruption, the resulting perturbations, and the propagation of these disturbances through the non-uniformly magnetized solar corona. By exploring how coronal waves produced by an eruption interacted with a distant flux rope prominence, we extended our previous studies \citet{Liakh:2020aap} and \citet{Liakh:2023aap}. Our main findings can be summarized as follows: 
\begin{itemize} 
        
        \item The EFR becomes unstable immediately after the experiment begins, generating a quasi-circular front. The EFR reaches approximately $100$ Mm in height when the dimming region of reduced density and temperature separating the EFR and the surrounding corona becomes visible. Below the EFR, the current sheet forms, which eventually fragments into plasmoids. The plasmoids form both above and below the reconnection null point, transporting hot, dense plasma to the EFR and post-reconnection loops at velocities exceeding $1000\kms$. The downward-moving plasmoids contain extremely hot plasma, with temperatures of $12$ MK.
        
        \item The EFR generates a front that propagates into the coronal medium, moving through regions with varying plasma-$\beta$ conditions.  %Fast-to-slow mode conversion arises when this wave crosses the equipartition line and forms a secondary front. 
        Synthetic SDO/AIA observations reveal that a fast EUV front propagates horizontally, initially reaching speeds above $1000\kms$ but then decelerating to $250\kms$. This front initially propagates as an ordinary fast magnetoacoustic wave but then transitions to a shock wave due to a decreased local phase speed. A slow secondary EUV front emerges as the fast EUV front traverses the equipartition lines. When the fast EUV front reaches the prominence region, crossing from the high to low $\beta$ region, it produces the reflected and transmitted EUV fronts. The fast EUV front propagates in the vertical direction, initially reaching speeds above $1000\kms$ and then decelerating to $600\kms$ due to the decreasing magnetic field strength with height. The slope of the primary front coincides with the one given by the propagation of the ordinary fast magnetoacoustic wave.      
           
        \item Following the formation of the flux rope, we load the prominence plasma into the magnetic dips. The prominence is affected by the compression toward the center of the magnetic dips, becoming a denser and narrower structure. Converging flows are observed toward the tail at the top of the prominence. Over time, these flows become evident throughout the rest of the prominence body. This process depletes the flux rope, accreting plasma onto the prominence body.
        
        \item The wave interacts with the flux rope prominence, inclining the flux rope prominence to the left, producing plasma motions around the centers of the magnetic dips. Additionally, the wave pushes the flux rope prominence down and triggers oscillations in the vertical direction and magnetic reconnection between the flux rope and post-reconnection loops. Analysis of fluid elements shows that these two types of oscillations have different characteristic periods and damping times, indicating distinct restoring and damping mechanisms.

\end{itemize}

%The comprehensive analysis of these effects enables us to construct a detailed understanding of EUV wave generation and propagation through the magnetized corona, including its interactions with distant prominences.
% 2.5D catastrophe eruption 
For the eruptive event, we used the setup from \citet{Takahashi:2017apj}, a 2.5D catastrophe scenario based on a force imbalance of the initial magnetic field. 
Despite the relative simplicity of this eruptive scenario, it successfully reproduces key features, including the shock front, dimming region, the compression layer surrounding the dimming region, the current sheet beneath the EFR, plasmoids, and post-reconnection loops. Compared to \citet{Zhao:2022apj}, we increased the strength of the cylindrical current by increasing its radius and obtained a more intense front due to the stronger magnetic field.
%

%Magnetic reconnection below the EFR
Perturbations below the EFR are initiated by the interaction between the reconnection outflow and the EFR. \citet{Takahashi:2017apj} showed that in similar configurations, upward-moving flows create a high-density envelope inside the EFR. We find that this perturbation propagates along the lateral sides of the EFR and converges, forming a region of high-density plasma that highlights the top part of the EFR.

%wavefront generation
Ahead of the EFR, the primary front is established. This front forms as a response to the initial force imbalance, generating the perturbation in the density and temperature distributions. This approach to producing coronal waves has been used in several numerical studies \citep[e.g.,][]{Wang:2009apj, Mei:2020mnras, Hu:2024apj}. The primary front becomes quasi-circular due to the anisotropic environment. Variations in magnetic field strength and plasma density along different directions influence how the perturbation propagates, thereby probing the surrounding coronal environment \citep{Liu:2019apj, Downs:2021apj}.

% Dimming region and compression front
A feature observed during the eruption is the formation of a dimming region characterized by reduced density and temperature, separating the expanding eruption from the surrounding corona. The decrease in plasma density and temperature within this region results in its dark appearance in synthetic and observed EUV images, as also reported by \citet{Attrill:2009apj}, \citet{Vanninathan:2018apj}, and \citet{Veronig:2019apj}, among others,  who identified similar dimming signatures associated with CMEs. In our synthetic images, the dimming region is surrounded by the compression layer, which is most evident in the 193 \AA\ channel, similar to the synthetic images shown in \citet{Downs:2011apj}. 

% Current sheet formation and plasmoids dynamics and properties
As the EFR rises, an elongated current sheet forms below. It fragments into plasmoids early in the experiment. By detecting and tracking the plasmoids inside the current sheet, we obtained their properties. Our numerical experiment suggests that the upward-propagating plasmoids accelerate to substantial velocities, exceeding $1000\kms$. At the same time, the downward-moving plasmoids have extremely high temperatures, around $12$ MK,  appearing bright in the 131 \AA\ synthetic images. Observations provided us with estimates for the properties of plasma blobs, which could be evidence of these plasmoids. Consistent with our results, high velocities of plasma blobs have been detected by \citet{Liu:2013mnras}. More recently, plasma properties in these plasma blobs have been estimated by \citet{Lu:2022apjl} and \citet{Hou:2024aap}. The temperatures and velocities estimates also agree with our findings. 
%Additionally, moderate velocities of the plasma blob have also been reported by \citet{Hou:2024aap}. Similarly, plasmoids with comparable properties were observed in our experiment.
%
In contrast to \citet{Zhao:2022apj}, we do not have plasmoids with chromospheric temperatures and densities, which originate from lifted and compressed chromospheric material in the current sheet before fragmentation. The reason is that we did not include the lower atmosphere in our initial atmosphere, but this can be addressed in future studies.
% However, the plasmoids demonstrate the transport of dense plasma upward to the EFR and downward to post-reconnection loops at very large speeds. The evolution of the current sheet, as well as the formation of the plasmoids and their properties, is likely defined by the magnetic field structure. It is interesting to investigate how this will evolve for the different parameters of the magnetic field structure. That is an interesting topic that can be addressed in future studies.
%  Generation of the wavefronts, fast, slow fronts, fronts from plasmoid motions, and appearance in the synthetic images.
%In our numerical experiment, the eruption produces fast and slow fronts detected in the $\nabla\cdot\mathbf{v}$ distribution. Similarly, as in \citet{Mei:2020mnras}, the fast front is evident, propagating far from the eruption, and the slow front is faint and hardly observed in synthetic images.
 The $\nabla\cdot\mathbf{v}$ distribution shows that multiple fronts propagate away from the eruption and are associated with fast-moving plasmoids in the current sheet. The perturbations produced in this case are important for driving small-amplitude oscillations of nearby solar prominences, as shown in our recent study \citep{Liakh:2023aap}. However, in the present case, these waves are fully damped before reaching our distant prominence.

 %Horizontal propagation of the fast front as a fast magnetoacoustic wave, the transition from magnetically- to gas-dominated regions, secondary front as a slow magnetoacoustic wave, evolution of the FMAW into a shock, the transition from gas-  to magnetically-dominated region, reflected front, secondary front as a SMAW—fast and slow EUV waves in our synthetic 193.
We studied the propagation of the eruption-generated primary front in both the horizontal and vertical directions. 
%The speed of the vertical propagation of the perturbation agrees well with the propagation of the fast magnetoacoustic wave, which decelerates propagating to a higher corona due to a decrease in the magnetic field strength. The synthetic 193 \AA\ images show the fast EUV wave, an ordinary fast magnetoacoustic wave.
 %
Horizontal propagation of the front is of particular interest. This resembles the behavior of a CME-induced perturbation traveling across the solar disk and interacting with local non-uniformities in the magnetic field and plasma density. In our numerical experiment, the magnetic field configuration consisted of three regions: the eruption, a smooth transition to the quiet Sun corona, and a transition to the prominence magnetic field. Our analysis reveals that, initially, the primary front propagates as an ordinary fast magnetoacoustic wave. When it reaches the equipartition layer, it produces the secondary front. Further analyzing the properties of this secondary front, we conclude that this is a slow magnetoacoustic wave similar to what has been obtained by \citet{Chen:2016solphys}. The formation of the secondary front has also been observed in regions of significantly reduced magnetic field strength \citep{Chandra:2022gal}. When the primary front moves even further into the region of the weak magnetic field, it becomes a shock wave as the local phase speed decreases significantly. The EUV wave as a weakly shocked fast magnetoacoustic wave has also been confirmed in observations \citep{Veronig:2010apjl, Takahashi:2015apj, Long:2015apj, Wang:2020apj}.

The primary front reaches the prominence region, producing a reflected front. The reflected fronts have been observed in the context of the interaction of EUV waves with different objects in the solar corona, such as coronal holes, loop systems, filaments \citep[see, e.g.,][]{Long:2008apjl, Gopalswamy:2009apjl, Li:2012apj, Shen:2012apj, Olmedo:2012apj, Kienreich:2013solphys,  Shen:2013apjl, Yang:2013apj, Chandra:2024apj}.  The primary front also produces a slower transmitted front that propagates along the magnetic loops and decelerates. The magnetic and acoustic flux analysis confirms that this is an ordinary slow magnetoacoustic wave. Moreover, the initial propagation speed agrees well with the sound speed. The most likely explanation is that this slow magnetoacoustic wave results from another fast-to-slow mode conversion. Our results agree with previous 3D simulations of the fast magnetoacoustic wave propagation from the gas-dominated region in the lower solar atmosphere to the magnetically dominated atmosphere \citep{Felipe:2010apj}. The conversion between fast and slow magnetoacoustic waves happens qualitatively in 3D, similar to what we see in two dimensions. The fast magnetoacoustic mode is transformed where $c_S = v_A$. After the transformation, a slow acoustic mode propagates along the field lines in the magnetically dominated region. The fast magnetic mode is reflected.

 %Vertical propagation of the fast front as a fast magnetoacoustic wave. Fast EUV front in synthetic 193.
%We studied the propagation of the primary front in the vertical directions. The speed of the vertical propagation of the perturbation agrees well with the propagation of the fast magnetoacoustic wave, which decelerates propagating to a higher corona due to a decrease in the magnetic field strength. The synthetic 193 \AA\ images show the fast EUV wave, an ordinary fast magnetoacoustic wave. The synthetic images also show the dark dimming region and bright compression layer due to the reconfiguration of the background coronal magnetic field due to the EFR rise. This agrees with results obtained in previous numerical studies \citep{Chen:2002apjl, Chen:2005apj, Cohen:2009apj, Downs:2011apj}.

%Thus, our numerical experiment and the corresponding synthetic images confirmed the formation of the dimming region and compression layer, the fast EUV wave, which was a fast magnetoacoustic shock wave produced by the eruption, the reflected EUV wave at the prominence region, and, finally, two slow EUV waves, which formed likely due to the mode conversion at the equipartition lines.

% Prominence interaction with the waves, comparison with observations and simulations
 Our experiment shows that the primary front pushes the prominence to the left and down. This triggers prominence dynamics around the centers of the magnetic dips and in the vertical direction. Analysis of the longitudinal and transverse velocities confirms the excitation of both oscillatory modes. In previous works \citep{Liakh:2020aap, Liakh:2023aap}, we obtained similar results using an artificial perturbation that created an energetic wave, simultaneously triggering longitudinal and transverse oscillations. Observations also showed the simultaneous driving of the different types of oscillations \citep{Gilbert:2008apj}.
 The longitudinal oscillations have smaller initial amplitudes, periods that vary with height, and longer damping times. 
 The transverse oscillations have a large initial amplitude that exceeds $14\kms$, a constant period at different heights, and a very short damping time. This means that these two types of oscillations have different restoring forces and damping mechanisms. 
 The variation of the period of longitudinal oscillations with height is well explained by the pendulum model \citep{Luna:2012apjl}. In this model, the main restoring force of the longitudinal oscillations is gravity projected along the magnetic field, and the period is defined by the radius of the curvature of the corresponding field line. In the 2.5D flux rope, the radius of the curvature decreases with height, resulting in a shorter period.
 The constant period of transverse oscillations with height was also found in our previous works \citep{Liakh:2020aap, Liakh:2023aap}, suggesting a global character of the motion. This period is well described by a model in which magnetic tension serves as the primary restoring force, which agrees with earlier works \citep{Zhou:2018apj, Adrover:2020aap}. 
 % 
%This period difference can lead to the so-called zig-zag shape of the prominence \citep{Luna:2016apj}. This is evident in the prominence evolution. 

Significantly different damping times for these two types of oscillations suggest that different mechanisms are involved. For longitudinal oscillations, several damping mechanisms can play an important role, such as non-adiabatic effects \citep{Zhang:2019apj, Fan:2020apj}, energy transfer across the field lines \citep{Liakh:2021aap}, mass accretion \citep{Ruderman:2016aap}, numerical dissipation \citep{Liakh:2020aap, Liakh:2021aap}, or wave leakage \citep{Zhang:2019apj}.  In our experiment, the plasma accretion on the main prominence body is observed. We estimated the damping time of the longitudinal oscillations following \citet{Ruderman:2016aap}. This damping time is $78.5\mins$, which greatly exceeds the one we obtained for these oscillations $20-40\mins$. Considering all of the above, the damping mechanism cannot be explained only by plasma accretion. However, this mechanism is important in addition to non-adiabatic effects, energy transfer, wave leakage, and numerical dissipation. 

For transverse oscillations, wave leakage has primarily been considered as a damping mechanism \citep{Zhang:2019apj, Liakh:2020aap, Liakh:2021aap}. In our experiment, mass accretion, particularly the accumulation of material at the top of the prominence, may also contribute to damping. As the total plasma mass increases, the average momentum decreases, which can attenuate the oscillations. Notably, the effect of mass accretion on the damping of transverse oscillations has not been thoroughly studied either analytically or numerically.

Another possible explanation for the significant damping of transverse oscillations is the magnetic reconnection triggered by the arrival of the primary front. This front pushes the flux rope downward, compressing its field lines against those of the post-reconnection loops, inducing reconnection. The current sheet shows an interesting evolution over this time. Initially, the vertical current sheet disappears, and the horizontal is formed instead. Over time, the current sheet realigns vertically again. Whether this process resembles oscillatory reconnection in the corona, as described by \citet{Karampelas:2022apj}, and whether it contributes to such significant damping remain open questions worth further investigation.

 Recent observations confirm that triggering of prominence oscillations by EUV waves can originate from eruptions \citep{Devi:2022adsr, ZhangY:2024apj, ZhangQM:2024mnras}. For instance, \citet{Devi:2022adsr} studied the EUV wave propagating with the speed $445\kms$. The authors detected the moment of interaction and pushing action of the EUV front on the prominence. After this interaction, the distant prominence showed transverse oscillations with periods in the range of $14-22\mins$. \citet{ZhangY:2024apj} investigated the transverse oscillations in the prominence and filament induced by the EUV wave with the average speeds $498\kms$ and $451\kms$, and obtained periods of $29.5-31.1\mins$, damping times of $44$ and $21\mins$. Finally, \citet{ZhangQM:2024mnras} observed one EUV wave with speed $835\kms$ and another decelerating from $788$ to $603\kms$. These waves induced prominence transverse oscillations that lasted only a few periods and were the most evident in the 304 \AA\ channel. The authors defined the main oscillatory characteristics, period and damping time, $18-27\mins$ and $33-108\mins$, respectively. Thus, our numerical experiment shows similar characteristics of the EUV waves and the induced prominences oscillations. We obtained a slightly shorter period for the transverse oscillations compared to observations. This discrepancy is associated with the relatively smaller size of the simulated prominence relative to the observed ones. According to our analysis, the prominence size is a key parameter influencing the period of transverse oscillations, assuming similar density and magnetic field strength. Therefore, incorporating larger prominences in our simulations would help reduce this difference.

\section{Conclusions}\label{sec:conclusions}
% Conclusions
In this study we investigated the interaction between eruptions and distant prominences through a 2.5D numerical experiment. Overall, this study provides a comprehensive view of the dynamics of coronal waves and prominence oscillations. We conclude that the fast EUV front produced by the eruption is a fast magnetoacoustic wave that evolves in the non-uniform solar corona. This evolution leads to the formation of a secondary EUV front, which can be interpreted as a slow magnetoacoustic wave. When the fast EUV wave encounters the flux rope prominence, it produces the reflected and transmitted EUV fronts. 
%Additionally, our numerical experiment and the corresponding synthetic images confirm the formation of the dimming region and compression layer.
The interaction between the fast EUV wave and the remote prominence is evident, and it manifests in the prominence oscillations and the triggering of magnetic reconnection at the null point below the prominence-hosting flux rope. The amplitudes, periods, and damping times of the prominence oscillations are in good agreement with previous numerical studies and observational results. However, our large-scale simulation is the first to combine these many facets in a single simulation that resolves details in all important regions, such as the current sheet, near the null point, or at the prominence location.

% Prospects, delayed eruption, realistic prominence formation, 3D configuration. 
Our results show the importance of studying the interaction between eruption-generated waves and prominences. Future work could explore how different magnetic field configurations allow delayed eruptions, as shown in \citet{Hu:2024apj}. Using a delayed eruption setup allows sufficient time for the prominence to form. We could also allow a more realistic scenario of prominence formation, such as by levitation-condensation or evaporation-condensation, bringing in the chromospheric layers. Additionally, understanding the effects of mass accretion on prominence oscillations requires further investigation, which would be more evident in the case of the evaporation-condensation scenario. 
This delayed eruption setup would also enable a reduction in the distance between the eruption and the prominence. Reducing the size of the numerical domain is particularly important for 3D simulations, as it significantly lowers computational costs. 
While a large domain was feasible in our 2.5D simulations due to AMR, such an approach is computationally impractical in 3D. The extension of this experiment to 3D is crucial because it allows us to study different orientations of the flux rope prominence structure with respect to the perturbing front. Furthermore, the anchoring of the prominence-hosting flux rope provides additional magnetic tension force that can significantly affect the triggering and properties of the transverse oscillations.

\begin{acknowledgements} 
We acknowledge support by the ERC Advanced Grant PROMINENT from the European Research Council (ERC) under the European Union’s Horizon 2020 research and innovation programme (grant agreement No. 833251 PROMINENT ERC-ADG 2018). RK acknowledges funding from Research Foundation Flanders (FWO) under grant agreement G0B9923N Helioskill, and under KU Leuven C1 project C16/24/010 UnderRadioSun. The computational resources and services used in this work were provided by the VSC (Flemish Supercomputer Center), funded by FWO and the Flemish Government, department EWI. We acknowledge using yt, an open-source analysis and visualization toolkit for astrophysical simulations, developed by the yt Project \citep{Turk:2011apjs}.
 \end{acknowledgements}

\bibliographystyle{aa} % style aa.bst
\bibliography{bibtex.bib}

\begin{appendix}
\FloatBarrier %\usepackage{placeins}
\onecolumn
\section{Additional figures} \label{sec:figures}
\begin{figure*}[!ht]
   \includegraphics[width=0.98\textwidth]{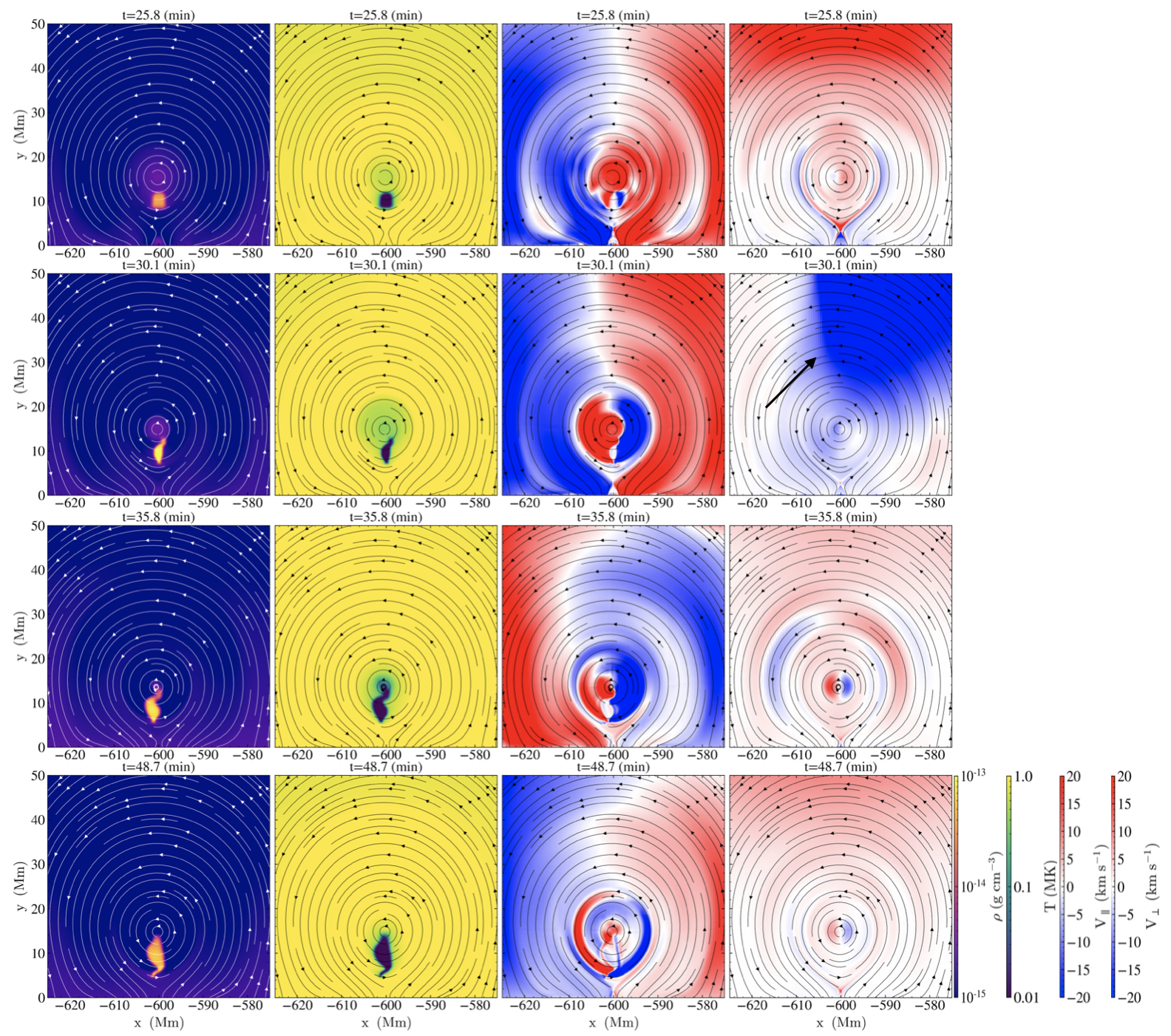}
    \caption{Density, temperature, $v_{\parallel}$, and $v_{\perp}$ distributions during various stages of the experiment, focusing especially on the deformed bipolar region that hosts the prominence: prominence mass loading (first row), the wave interaction with the prominence (second row), prominence oscillations and mass accretion (third and fourth rows). The black arrow denotes the primary front. Animation 4 shows the temporal evolution of the density, temperature, and the corresponding velocities up to $57.2\mins$. An animation of this figure is available online.}
    \label{fig:prominence_evolution}
\end{figure*}
\end{appendix}

\end{document}